\documentclass[fleqn,usenatbib]{mnras}

\usepackage{newtxtext,newtxmath}

\usepackage[T1]{fontenc}
\usepackage{ae,aecompl}

\usepackage{graphicx}	
\usepackage{amsmath}	
\usepackage{amssymb}	
\usepackage{multirow}
\usepackage{cuted}
\usepackage{xcolor}

\newcommand{\hda}{H$\delta_{\rm A}$}
\newcommand{\hga}{H$\gamma_{\rm A}$}
\newcommand{\ha}{H$\alpha$}
\newcommand{\sfrunit}{\ensuremath{M_\odot/yr}} 
\newcommand{\themerger}{2xSc\_07 } 
 \newcommand\orcidicon[1]{\href{https://orcid.org/#1}{\includegraphics[scale=0.65]{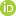}}}
\definecolor{orcidlogocol}{HTML}{A6CE39}

\title[Stellar populations in simulated PSBs]{Comparison of stellar populations in simulated and real post-starburst galaxies in MaNGA}

\author[Y. Zheng et al.]{
Yirui Zheng\orcidicon{0000-0001-7707-5930},$^{1}$\thanks{E-mail:yz69@st-andrews.ac.uk} 
Vivienne Wild\orcidicon{0000-0002-8956-7024},$^{1}$
Natalia Lah\'en,$^{2}$
Peter H. Johansson\orcidicon{0000-0001-8741-8263},$^{2}$
David Law,$^{3}$
\newauthor
John R. Weaver\orcidicon{0000-0003-1614-196X},$^{4,5,1}$
Noelia Jimenez$^{1}$
\\
$^{1}$School of Physics and Astronomy, University of St Andrews, North Haugh, St Andrews, Fife, KY16 9SS, Scotland, UK\\
$^{2}$Department of Physics, University of Helsinki, Gustaf 
H$\ddot{a}$llstr$\ddot{o}$min katu 2a, FI-00014 Helsinki, Finland, \\
$^{3}$Space Telescope Science Institute, 3700 San Martin Drive,
Baltimore, MD 21218, USA\\
$^{4}$Cosmic Dawn Center (DAWN)\\
$^{5}$ Niels Bohr Institute, University of Copenhagen, Lyngbyvej 2, Copenhagen \O~2100, Denmark
}

\date{Accepted XXX. Received YYY; in original form ZZZ}

\pubyear{2020}

\begin{document}
\label{firstpage}
\pagerange{\pageref{firstpage}--\pageref{lastpage}}
\maketitle

\begin{abstract}

Recent integral field spectroscopic (IFS) surveys have revealed radial gradients in  the optical spectral indices of post-starburst galaxies, which can be used to constrain their formation histories. We study the spectral indices of post-processed mock IFS datacubes of binary merger simulations, carefully matched to the properties of the MaNGA IFS survey, with a variety of black hole feedback models, progenitor galaxies, orbits and mass ratios. Based on our simulation sample, we find that only major mergers on prograde-prograde or retrograde-prograde orbits in combination with a mechanical black hole feedback model can form galaxies with weak enough ongoing star formation, and therefore absent \ha\ emission, to be selected by traditional PSB selection methods. We find strong fluctuations in nebular emission line strengths, even within the PSB phase, suggesting that \ha\ selected PSBs are only a subsample of the underlying population. The global PSB population can be more robustly identified using   
stellar continuum-based approaches. The difficulty in reproducing the very young PSBs in simulations potentially indicates that new sub-resolution star formation recipes are required to properly model the process of star formation quenching. 
In our simulations, we find that the starburst peaks at the same time at all radii, but is stronger and more prolonged in the inner regions.
This results in a strong time evolution in the radial gradients of the spectral indices which can be used to estimate the age of the starburst without reliance on detailed star formation histories from spectral synthesis models.

\end{abstract}




\section{Introduction}
\label{sec:intro}
The galaxy population in the local Universe shows a clear bimodality in the
colour-magnitude diagram \citep{Baldry_2004, Jin_2014}. 
The majority of massive galaxies fall into two distinct groups:
the ``blue cloud''  (star-forming, spiral galaxies) and the ``red sequence''  (quiescent, elliptical galaxies). The total mass of the stars living in quiescent galaxies, as well as the total number of quiescent galaxies, has grown steadily since a redshift of at least $z\sim4$ \citep{Ilbert2013,Muzzin2013}, which implies a steady conversion of star-forming to quiescent galaxies through the quenching of their star formation. 

There may be multiple evolutionary processes by which galaxies migrate from the blue cloud  to the red sequence. Analysis of the star formation histories (SFHs) of quiescent galaxies has revealed at least two different quenching mechanisms that operate on different timescales, ``fast''  and ``slow'' quenching. At high redshift, galaxies of all stellar masses consistently show a fast rise and decline of their star formation rate (SFR). At low redshift, low-mass galaxies grow slowly and quench quickly, whereas high-mass galaxies grow quickly and quench slowly \citep[e.g.][]{Pacifici2016}.

Post-starburst (PSB) galaxies, also sometimes referred to as E+A or K+A galaxies, may occupy an important transitional state between the star-forming and quiescent populations, and could be an important candidate for the ``fast'' quenching route. These galaxies have experienced a rapid decline in their star formation following a previous period of very rapid star formation. Catching these galaxies as they transition to quiescence may provide important constraints  on the causes of the rapid quenching which is not possible from archaeological studies of already-quiescent galaxies. They are identified by their excess population of intermediate-age (A- or F-type stars), but deficit or absence of hotter, younger stars (O- and B- types) leading to strong Balmer absorption lines, a strong Balmer break, and weak or absent nebular emission lines from star formation. PSBs account for only about 1 per cent of local galaxies \citep{goto2008integrated, wong2012galaxy}, however they may account for a significantly larger fraction of the galaxy population at higher redshifts \citep{Tran2004,wild2009post,vergani2010,Whitaker2012,wild2016,Rowlands2018GAMA,Belli2019}. Despite being rare, their number density is sufficient to account for a significant fraction of the growth of the red-sequence at $z<2$, although the precise fraction is still debated  \citep{Tran2004,wild2009post,wild2010timing,Belli2019}.

At high redshift, the morphologies of PSB galaxies are strongly suggestive of an extremely intense and rapid quenching process, such as a major gas-rich merger, multiple mergers or protogalactic collapse. \citet{Almaini2017} find that massive PSBs at redshift ($z>1$) are significantly smaller than comparable quiescent galaxies at the same stellar mass and epoch which is strongly suggestive of a recent dissipative collapse event associated with the shut off in star formation \citep[see also][]{Yano2016}. This appears to depend on stellar mass and epoch, with lower mass and lower redshift PSBs more similar in structure to equivalent quiescent galaxies \citep{Maltby2018, pawlik2018origins}.

At lower redshift, galaxy mergers have long been proposed as a possible formation mechanism for PSBs shortly after the discovery of  E+A galaxies \citep{Lavery1988}. 
A substantial fraction of PSBs are found with faint tidal features or companion galaxies \citep{zabludoff1996environment, Chang2001, pawlik2018origins}. In addition, PSBs at low-redshift can have surprisingly high fractions of mass formed in the starburst of typically 20\%, but occasionally as much as 70\% \citep{Kaviraj2007, pawlik2018origins}, which again supports major mergers as the only mechanism known in the local Universe to be able to feed sufficient gas into the central regions of a galaxy in a short enough time \citep[see also the discussion in][]{Weaver2018}. Such a scenario is consistent with the results from the cosmological hydrodynamic simulation EAGLE, where \citet{davis2019evolution} 
find that local ($z\sim0$) simulated post-starburst galaxies are predominantly caused by major mergers. \citet[][]{pawlik2019} used the same EAGLE simulations, post-processed to produce mock optical spectra, to find that in addition to classical major mergers, a range of different processes can create weak PSB features detectable in modern spectroscopic surveys. These include harassment by multiple smaller galaxies as well as rejuvenation from gas brought in by infalling small satellites. 

During a galaxy merger, the discs of the galaxies are typically destroyed and the gas components are funnelled into the galaxy centre, leading to a powerful centralised starburst followed by rapid quenching of the star formation \citep[e.g.][]{barnes1992transformations, naab2003statistical, bournaud2005galaxy}. However, this induced starburst may not cause sufficient quenching to produce remnants with no ongoing star formation, as without a mechanism to expel the gas directly, the remnant typically continues to form stars \citep[e.g.][]{Sanders1988, hopkins2006unified, Johansson2009equal}. The more or less complete prevention of star formation in quiescent galaxies is the primary motivation for black hole feedback in the current generation of galaxy evolution simulations. In this model, an active galactic nucleus (AGN) can heat up the cold gas in the galaxies by thermal feedback and/or expel the cold gas by mechanical feedback, resulting in a lack of material for further star formation. Several studies suggest that PSBs may harbour an excess fraction of AGN compared to normal galaxies \citep[e.g.][]{yan2006,wild2007bursty, wild2010timing, pawlik2018origins}, however the short duty cycles of AGN, compared to the observed post-starburst features, hamper any direct comparison. It is also possible that external conditions play a role in the ultimate quenching of post-merger remnants, which may explain why PSB galaxies are more common in (some) clusters and intermediate density environments \citep[e.g.][]{zabludoff1996environment, Poggianti2009, Socolovsky2018, pawlik2018origins, Paccagnella2019}. Ram-pressure stripping, strangulation or harassment may all play a role in suppressing further star formation in merger remnants that find themselves in dense environments. 

The development of spectral synthesis models allowed the first theoretical explorations of the unusual spectral properties of PSB galaxies \citep[e.g.][]{couch1987, poggianti1999star, shioya2004formation} leading to a basic understanding of their star formation histories. However, significant progress on understanding the formation mechanisms was only made once the dynamics of the gas in galaxies could be tracked in detail. Early hydrodynamic gas-rich merger simulations combined with simple star formation laws showed that gas rich mergers could lead to star formation histories characterized by strong, short bursts \citep{Mihos1994b,Mihos1996}. Combining hydrodynamical galaxy merger simulations with spectral synthesis models, led \citet{Bekki2005} to reproduce the positive colour gradient and negative radial H$\delta$ gradient observed in local PSBs by \citet{pracy2005}. A similar analysis was carried out by \citet{wild2009post}, at higher resolution and using more sophisticated star formation and feedback models, including a comparison between models with and without BH feedback. They also found that the stellar continuum properties of PSB galaxies at $0.5<z<1$ could be reproduced by merger simulations with starburst mass fractions larger than $\sim5-10$\% and decay times shorter than $\sim10^8$\,yr. BH feedback was not required to reproduce the stellar continuum features of PSB galaxies, although they did not investigate the emission line properties. Similar simulations were carried out by \citet{snyder2011k+}, who again concluded that the role of AGN feedback in ceasing star formation and producing PSB features was negligible. Since these studies were undertaken the sub-resolution recipes used in the hydrodynamic simulations have become much more sophisticated and the spatial resolution has increased, motivating a revisit of the question of how the unusual spectral properties of post-starburst galaxies can be reproduced. 

The recent advent of large integral field spectroscopic (IFS) surveys of local galaxies, including CALIFA \citep{Sanchez2012}, SAMI \citep{Croom2012}, and MaNGA \citep[Mapping Nearby Galaxies at Apache Point Observatory;][]{MaNGA_overview}, has led to greater interest in the spatial properties of galaxies. In this paper, we revisit the question of radial gradients in the spectral properties of MaNGA post-starburst galaxies \citep{Chen2019}, using up to date simulations to investigate the quenching mechanisms at play in these galaxies.  

The outline of the paper is as follows: in Section \ref{sec:observations} we motivate our analysis by presenting the radial gradients in spectral indices observed in three example MaNGA galaxies; in Section \ref{sec:simulation} we introduce the merger simulations; in Section \ref{sec:sedmorph} we describe how we create mock MaNGA datacubes for the simulated galaxies; we present our analysis of the mock PSBs in Section \ref{sec:results}, including their global and radial gradient properties; finally in Sections \ref{sec:Discussion} and \ref{sec:Summary} we discuss and summarise our results.

\section{Observational Motivation}
\label{sec:observations}

\begin{figure}
	\includegraphics[width=0.95\columnwidth]{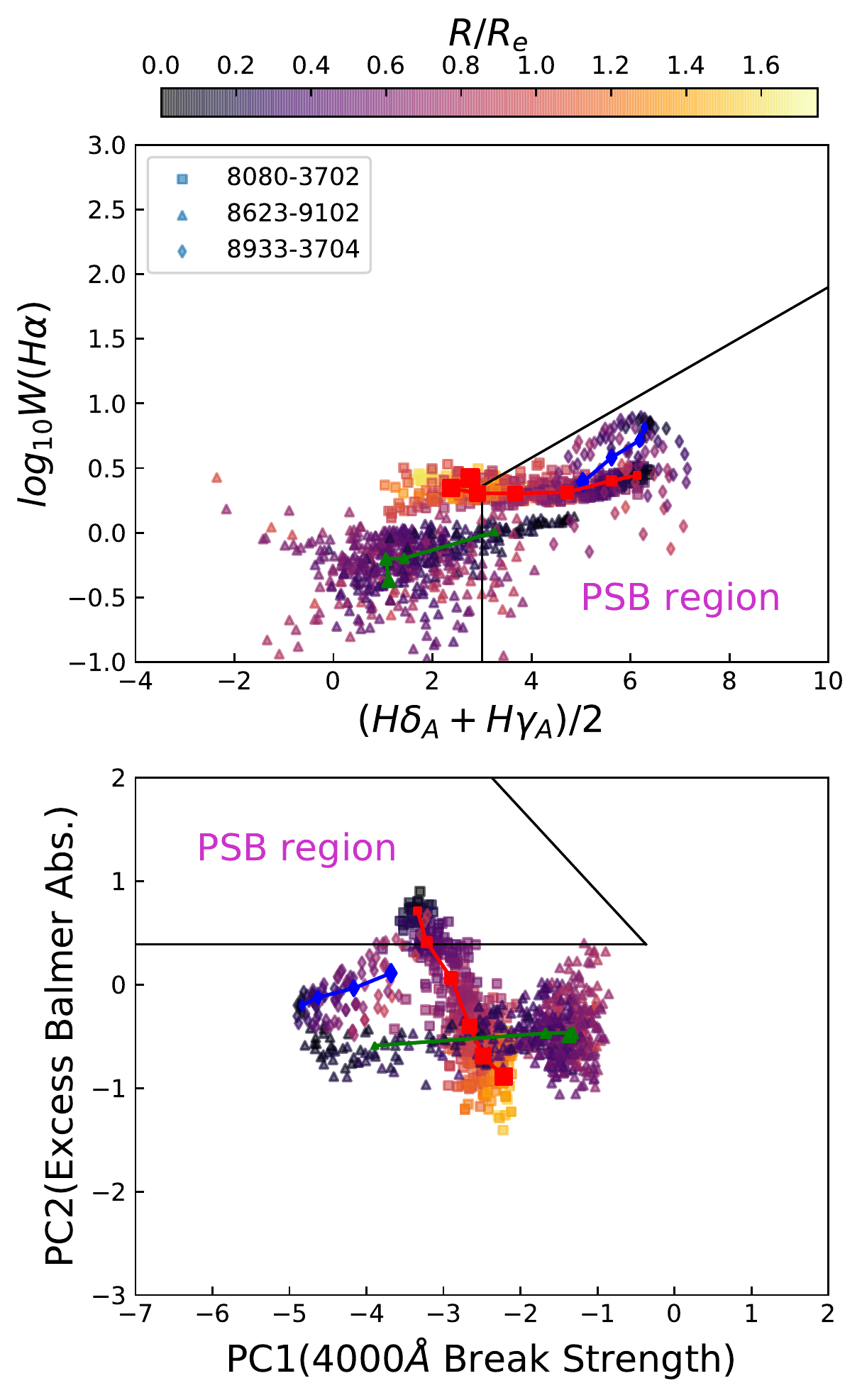}
	\caption{Spectral index distributions for three representative MaNGA post-starburst galaxies selected from \citet{Chen2019}, with points colour coded by radius with respect to the galaxy's effective radius as given by the upper colour bar. The red, green and blue lines and points show the median values for spaxels binned in annuli of width 0.25\,R$_e$ with larger symbols indicating outer regions. The MaNGA plate-ifu identifiers are given in the inset box. \textit{Top:} \ha\ equivalent width (in \AA) vs. the summed stellar continuum indices \hda\ and \hga. \textit{Bottom:} the principal component spectral indices that characterise the strength of the 4000\AA\ break (PC1) and any excess Balmer absorption over that expected for the normal star-forming main sequence (PC2). 
	}
	\label{fig:observations}
\end{figure}

Traditionally, post-starburst galaxies are selected based on strong Balmer absorption lines, typically H$\delta$, with weak or absent nebular emission lines, commonly either the H$\alpha$ emission line \citep[e.g.][]{goto2003hdelta,quintero2004selection,balogh2005near}
or the [OII] line if the H$\alpha$ line is unavailable, as is common in higher redshift observations
\citep[e.g.][]{dressler1983spectroscopy, zabludoff1996environment, poggianti1999star, Maltby2016}.
However, selection based on nebular emission will result in incomplete samples, as narrow line nebular emission can also be caused by shocks and AGN, both of which appear to be prevalent in PSBs \citep{yan2006,wild2007bursty,wild2010timing,Alatalo2016, pawlik2018origins}. Additionally, in order to give us a complete picture of the galaxy quenching mechanisms we may be interested in objects that have had a recent starburst, but that have not (yet) completely quenched their star formation. More recently, alternative methods to select PSBs have been developed, that relies on the stellar continuum shape alone. \citet[][]{wild2007bursty} developed a principal component analysis (PCA) method to select PSBs from high quality spectral observations of the stellar continuum around 4000\AA. Similarly, photometric identification is possible in the case of the strongest spectral features \citep{Whitaker2012,Wild2014b, Forrest2018}. 

\citet{Chen2019} identified $>300$ galaxies with PSB regions based on traditional cuts on Balmer absorption line and \ha\ emission line strengths, out of over 4000 galaxies in the MaNGA Product Launch 6 \citep[MPL6, see][for an overview of the MaNGA survey]{MaNGA_overview}\footnote{An internal team data release including the first 3 years of MaNGA survey data}. Of these, they identified 31 galaxies with centralised PSB signatures (CPSB). For these 31 galaxies, we downloaded the publicly available spectral data cubes and spectral measurement maps from the SDSS Data Release 15 \citep[DR15,][]{MaNGA_DR15}, 
which were created by the MaNGA Data Reduction \citep[DRP, ][]{Law2016} and Analysis \citep[DAP, ][]{MaNGA_DAP} pipelines.
In both cases we used the ``HYB10'' spaxel binning where the stellar-continuum analysis is performed on Voronoi binned cells with a minimum signal-to-noise ratio (SNR) of 10 \citep{voronoi}, while the emission-line and spectral-index measurements are performed on the individual spaxels. 

From the DAP maps, we extracted the \ha\ emission line equivalent width (W(H$\alpha$)), \hda\ and \hga\ stellar continuum absorption line indices \citep{HdA}, the mean g-band weighted signal-to-noise ratio per pixel map and stellar velocity map.  In all that follows we masked spaxels with spaxel $g$-band SNR smaller than 10. During the MaNGA data analysis the emission and absorption lines are fit simultaneously, and the emission component is subtracted from the spectrum before the absorption indices are calculated \citep[see ][for full details]{MaNGA_DAP_emissionlines}. In this way the stellar and gas components of the spectrum are separated and are therefore independent of one another.  The W(H$\alpha$) measurement is then calculated from a Gaussian fit to the residual emission line spectrum.  The \hda\ and \hga\ absorption line indices are defined as the continuum normalised total flux difference between the absorption feature and a ``pseudo-continuum'' defined as a straight line between two flanking bandpasses, with the bandpasses as defined in \citet{HdA}. They can therefore be negative in old stellar populations whose spectra exhibit only weak Balmer absorption.

From the summary ``DRPall'' file we obtained the effective radius, measured as the radius that contains half of the Petrosian flux in $r$-band elliptical apertures from the NASA-Sloan Atlas catalogue\footnote{\url{http://nsatlas.org}}, and the Galactic extinction. From the DAP spectral data cubes we obtained the flux and variance, which we corrected for Galactic extinction and shifted to the rest frame using the redshift and stellar velocity maps. We then linearly interpolated the flux and error arrays onto the eigenbasis wavelength array of \citet[][]{wild2007bursty}, in order to measure principal component based spectral indices that describe the shape of the 4000\AA\ break region of the optical stellar continuum spectrum, using a normalised-gappy PCA algorithm to take account of the error arrays and any masked spaxels. Further details on the PCA spectral indices are given in Section~\ref{subsec:PCA}.

Following the method of \citet[][]{Chen2019}, we mask all spaxels with SNR$<10$ and those with ``DONOTUSE'' flags, and calculate the distance of every spaxel from the centre of the galaxy, as a function of the effective radius. In \autoref{fig:observations} we show the spectral index distribution as a function of distance from the centre for three good quality CPSB galaxies, in both W(H$\alpha$) vs. Balmer absorption line strength (top), and PC1 vs. PC2 (bottom). These three galaxies were selected to illustrate the range in radial gradients observed in the full sample. By definition, the central regions of the three galaxies lie within the traditional W(H$\alpha$) vs. Balmer absorption line strength selection box, which we define as (\hda+\hga)$/2 > 3$ and W(\ha) $< 2.2\times$(\hda$+$\hga)$/2 - 0.3$. We clearly see strong radial gradients in the Balmer absorption line strength, with the central regions much stronger than the outer regions. In the PCA spectral indices we see a range of radial gradients, from marginally positive to strongly negative. We note that neither 8933-3704 nor 8623-9102 would be identified as a PSB by the PCA spectral indices, within the selection box defined by \citet{Rowlands2018SDSS}, which suggests that a very recent and rapid shut down in their star formation has occurred, before the Balmer absorption lines have had a chance to strengthen sufficiently to reach the PCA selection box. 

Simple toy models indicate that these observations are consistent with two simple scenarios: a single co-eval burst which was stronger in the central regions, or a starburst that has progressed from outside-in (Weaver et al. in prep). However, the toy models are unable to distinguish between the two options. These results form the motivation for this paper: why do we observe radial gradients in the spectral indices of PSB galaxies, and can we use them to tell us something about the physical processes that led to the formation of these unusual galaxies?

\section{Simulations}
\label{sec:simulation} 
Our binary merger simulations are run with the N-body smoothed particle hydrodynamics  (SPH) code, SPHGal \citep{hu2014sphgal,eisenreich2017active}, which is an updated version of the Gadget-3 code \citep{springel2005cosmological}. Compared with the original Gadget-3 code, 
SPHGal replaces the spline kernel with a Wendland $C^4$ kernel and increases the number of neighbours in the SPH kernel to 100.
The code employs a set of new features including the pressure-entropy formulation of SPH, an updated estimate of velocity gradients, a modified artificial viscosity switch with a modified strong limiter, artificial conduction of thermal energy  and a time step limiter \citep[see][for more details]{lahen2018fate}.
Together these changes reduce  numerical artefacts in the fluid mixing and improve the convergence rate of the SPH calculation.

\subsection{Stellar sub-resolution physics}
\label{subsec:subgrid}
The subresolution astrophysics models in SPHGal are based on those by \citet{Scannapieco2005, Scannapieco2006}, 
  updated by \citet{Aumer2013}, 
 and include gas cooling, star formation, chemical evolution and stellar feedback. In this model, the gas component cools with a rate dependent on its temperature, density and metal abundance. Assuming that the gas is optically thin and in ionisation equilibrium, the cooling rates are calculated following \citet{Wiersma2009}, 
 on an element-by-element basis. The effects of a uniform redshift-dependent ionising UV/X-ray background \citep{haardt2001modelling} are also included assuming $z = 0$. The cooling rates are calculated over a temperature range of $10^2 \leq T \leq 10^9\ K$.

Dense and cold gas particles are able to form stars, once the gas density $\rho_g$ is greater than $\rho_{crit}=1.6\times 10^{-23} \ \rm g/cm^3$,  i.e  $n_H=10 \ \rm cm^{-3}$, and the gas temperature is less than $T=12000 \ \rm K$. The probability that a gas particle converts into a stellar particle is $1-e^{-p}$,
where p is given by:
\begin{equation}
    p=\epsilon_{SFR}\frac{\Delta t}{t_{dyn}} = \epsilon _{SFR} \Delta t \sqrt{4 \pi G \rho_g}.
\end{equation}
$\Delta t$ is the length of the current time step,  $t_{dyn}$ is the local dynamical time, and $\epsilon_{SFR}$ is the star formation efficiency. We adopt a fixed efficiency of $\epsilon_{SFR} =0.02$. 

To track the chemical evolution in the simulations, every baryonic particle contains 11 elements  (H, He, C, Mg, O, Fe, Si, N, Ne, S, Ca) that evolve based on models of chemical release rates from \citet{Iwamoto1999} 
for supernovae type Ia  (SNIa), 
\citet{Woosley1995} 
for supernovae type II  (SNII) and \citet{karakas2010updated} for asymptotic giant branch  (AGB) stars. The star particles distribute the metals to the surrounding gas particles by the stellar feedback of SNIa, SNII and AGB stars. The metallicity diffusion implementation by \citet{Aumer2013}
is included here to smooth the variations in the metallicity between neighbouring gas particles.

The stellar particles provide feedback the surrounding gas via SNII 3\,Myr after their formation, followed by the SNIa for which feedback is released repeatedly every 50 Myr from a stellar age of 50\,Myr until 10\,Gyr, with the assumption that the ejecta mass decays proportionally to $t^{-1}$  \citet{maoz2011nearby} and the total release is 2 SNIa per $1000 M\odot$ of stellar mass. The total feedback energy by SN to the interstellar matter  (ISM) is given by
\begin{equation}
    E_{SN} = \frac{1}{2}m_{eject}v_{SN}^2
\end{equation}
where $m_{eject}$ is the mass of SN ejecta, $v_{SN}$ is the velocity of SN ejecta  (we adopt $v_{SN} = 4000\ {\rm km/s}$ here) and the SN ejecta in our model is metallicity dependent.

The thermal and kinetic feedback of SN are achieved by a distance dependent multiphase method. The phases here are the free expansion  (FE) phase, the adiabatic SedovTaylor  (ST) phase \citep{taylor1950formation, Sedov1959} 
and the snowplow  (SP) phase  \citep{mckee1977theory, blondin1998transition}.
The gas particles that are closest to the SN receive the feedback in the FE phase with momentum conservation and the feedback energy is in kinetic form only. At greater distances, the shocked ISM mass exceeds the SN ejecta mass, gas particles in this outer region then receive feedback in the ST phase with heating. In the ST phase, the feedback energy are in both thermal  (70\%) and kinetic  (30\%) forms. At even larger distances, the velocity of the SN ejecta decreases further, finally dispersing the shock. Gas particles in this outermost region receive feedback in the SP phase with efficient radiative cooling. Again both thermal  (70\%) and kinetic  (30\%) energy is assumed, however the total amount of energy decreases with distance from the SN.

Both the energy and enrichment feedback from AGB stars are dealt with in the same fashion as that from the type Ia SNe. However, as the wind velocity of AGB ejecta is just $25$ km/s, much smaller than $v_{SN}$, only the FE phase is included for AGB feedback.

\subsection{Thermal black hole feedback model}
\label{subsec:oldBH}
We run a set of merger simulations without black hole  (BH) feedback and find that extra centrally injected energy is required in order to suppress the star formation in the merger remnants sufficiently to obtain the small \ha\ equivalent widths observed in many post-starburst galaxies. We investigate two different black hole models: a ``classical'' model with thermal energy injection only  \citep{Springel2005BHmodel, Johansson2009equal}, and an ``updated'' model with both thermal and mechanical energy injection \citep{Choi2012}.

In the classic model the unresolved gas accretion onto the BH is parameterised by the Bondi–Hoyle–Lyttleton model  
\citep{hoyle1939effect, bondi1944mechanism, bondi1952spherically}, in which the BH accretion rate is given by:
\begin{equation}
\label{eq:MB}
    \dot{M}_B=\frac{4\pi \alpha M_{BH}^2\rho}{ (c_s^2+v^2)^{3/2}}
\end{equation}
where $\rho$ is the gas density, $c_s$ is the sound speed of the surrounding gas, $v$ is the velocity of the BH relative to the surrounding gas, and $\alpha$ is a dimensionless efficiency parameter set to enable the self-regulation of the BH mass growth and ensure that the BH accretion reaches the Eddington regime in a gas-rich environment \citep{Johansson2009equal}. Physically, we should have $\alpha=1$, however \citet{Springel2005BHmodel} and \citet{Johansson2009equal} use a value of $\alpha = 100$ to account for the underestimated gas density or the overestimated gas temperature near the Bondi radius. As the spatial resolution of the simulations is increased, the local gas density and temperature is more accurately modelled, and $\alpha$ must be decreased. Here we set $\alpha = 25$ and confirmed that this value allows the BHs in the merger remnants to grow onto, and stay on, the $M_{BH}-\sigma$ relation \citep{Bosch2016}.

The accretion rate is limited to the Eddington accretion rate:
\begin{equation}
    \dot{M}_{edd}=\frac{4\pi G M_{BH} m_p}{\epsilon_r \sigma_T c}
\end{equation}
where $m_p$ is the proton mass, $\sigma_T$ is the Thomson cross-section and $\epsilon_r$ is the radiative efficiency. Here we adopt $\epsilon_r =0.1$, the mean value of radiatively efficient accretion onto a Schwarzschild BH  \citep{shakura1973black}. Thus, the final  (inflowing) accretion rate is given by:
\begin{equation}
	\label{eq:Minf}
    \dot{M}_{in}=min (\dot{M}_B, \ \dot{M}_{edd}).
\end{equation}

The radiated luminosity of the BH is given by:
\begin{equation}
    L_r=\epsilon_r \dot{M}_{in} c^2.
\end{equation}
We assume that a fraction, $\epsilon_f$, of radiated energy couples with the surrounding gas leading to the feedback energy: 
\begin{equation}
    E_{\rm feed}=\epsilon_f L_r=\epsilon_f \epsilon_r \dot{M}_{in} c^2. 
\end{equation}
We adopt a fixed value of $\epsilon_f = 0.05 $, a value that is widely chosen in the literature
\citet[]{Springel2005BHmodel, DiMatteo2005a, Johansson2009equal,Johansson2009evo,  Choi2012}.
The choice of this value for $\epsilon_f$ is motivated by the fact that in combination with the thermal feedback model the simulations are able to 
reproduce the observed $M_{BH}-\sigma$ relation. 

In this classical thermal black hole feedback model, all the feedback energy is distributed as thermal energy into the $\sim100$ gas particles closest to the BH, weighted by the SPH kernel. The gas around the BH is therefore heated by the BH feedback and expands, reducing its density, which leads to a cessation in star formation.

In our merger simulations, we assume that the binary BHs merge as soon as their separation drops below the smoothing length and their relative velocity drops below the local sound speed of the surrounding gas. Due to the limited spatial resolution the BHs can wander away from the centre of the galaxies, especially in unequal-mass mergers. To ensure the successful merging of the BHs during the final coalescence of their host galaxies, at every time step we re-locate the BH to the minimum potential in the central region of the galaxy 
\citep[see][for more discussion on this procedure]{Johansson2009equal}.

\subsection{Mechanical black hole feedback model}
\label{subsec:newBH}
It has been suggested that a strong wind from an accreting BH might convey energy, mass and momentum from the centre to the surrounding gas \citep{deKool2001keck, moe2009quasar, dunn2010quasar}, causing the gas outflows and the regulation of star formation in the host galaxies \citep{Tremonti2007ApJ...663L..77T, Feruglio2010A&A...518L.155F}. To simulate this effect we create a third set of simulations in which we adopt the mechanical BH feedback model developed by  Choi et al. (\citeyear{Choi2012, Choi2014}). 

In this model the total accretion rate onto the BH is given by:
\begin{equation}
	\label{eq:Macc}
    \dot{M}_{acc}= \dot{M}_{in} - \dot{M}_{out}
\end{equation}
where $\dot{M}_{out}$ is the outflowing mass loss rate, and $\dot{M}_{in}$ is calculated as above  (Eqn.~\ref{eq:Minf}) with $\dot{M}_B$ computed with an `alternative averaging (AA)' method \citep[see][]{Choi2012}. 
We further define the kinetic energy rate of the outflowing wind: 
\begin{equation}
	\label{eq:Ew}
    \dot{E}_{w}= \epsilon_f \epsilon_r \dot{M}_{acc} c^2  = \frac{1}{2}\dot{M}_{out}v_w^2
\end{equation}
where $\epsilon_r$ and $\epsilon_f$ are, respectively, the radiative efficiency and fraction of the radiated energy that couples to the surrounding gas  (as above), and $v_w$ is the velocity of the wind assuming energy conservation and a single wind velocity. 

Following \citet{Choi2012} we define the ratio of the mass outflow rate to the mass accretion rate as:
\begin{equation}
\label{eq:psi}
\psi \equiv \dot{M}_{out}/\dot{M}_{acc} = 2\epsilon_f \epsilon_r  c^2/v_w^2.
\end{equation}
Since the momentum is conserved, the total accretion rate $\dot{M}_{acc}$ and the kinetic energy rate $\dot{E}_w$ can be solved out as: 
\begin{subequations}
\label{eq:solution}
\begin{equation}\dot{M}_{acc}= \dot{M}_{in}\frac{1}{1+\psi}\end{equation}
\begin{equation}\dot{E}_w=\epsilon_f \epsilon_r c^2 \dot{M}_{in}\frac{1}{1+\psi} \end{equation}
\end{subequations}
Inspired by observations of broad absorption line winds 
\citep{crenshaw2003mass, moe2009quasar},
we assume a wind velocity of $v_w = 10,000\,{\rm km/s}$. With the widely adopted feedback efficiencies of $\epsilon_f \epsilon_r = 0.1 \times 0.05 =0.005$, Eqn.~\ref{eq:psi} gives $\psi=0.9$, i.e., 90\% of the inflowing mass is ejected in an AGN wind, with both energy and mass carried from the BH to the outskirts of the galaxy. This outflowing gas collides with the ambient ISM on its way out, resulting in a momentum-driven flow. We simulate this phenomenon by allowing the emitted wind particle to share its momentum with its 2 nearest neighbouring gas particles. As the gas particles all have the same mass in our simulation, all three gas particles gain the same velocity increment of $\Delta v \sim (v_w/3)\ {\rm km/s}$. This treatment conserves the momentum but decreases the total kinetic energy. To conserve the total energy, the residual energy is deposited into these three particles in form of thermal energy.

\citet{Choi2012} include an additional thermal feedback component in their simulations to simulate the effect of X-ray radiation from the AGN on the surrounding gas. However, we found that this was too effective at suppressing star formation, eliminating the starburst during the merger phase and preventing the appearance of the strong Balmer absorption lines seen in PSB galaxies. We therefore only ran one simulation with this effect included, which is presented in Section \ref{sec:results} below. We note that the updated stellar feedback model used in this paper, compared to the stellar feedback model used in  \citet{Choi2012}, may in part be responsible for the dramatic suppression of star formation seen in our simulations when including this additional thermal X-ray BH feedback component. 

\subsection{Galaxy models}
\label{subsec:galaxy}

\begin{table}
 \caption{Parameters for the progenitor disc galaxies. Total masses, $M$, are given in units of $10^{10} M_\odot$, scale radii and heights are in kpc, $N$ is particle number, $m$ gives the individual particle masses, $\epsilon$ the gravitational softening lengths and SFR is in M$_\odot$/yr. Other parameters are defined in the text.}
 \label{tab:ic_parameters}
 \centering
\begin{tabular}{lcccc}
\hline\hline
Property & Sa & Sc & Sd & Scp3\\
\hline
$M_{vir}$ & 134.1 & 134.1 & 134.1 & 44.7 \\
$M_{disc, *}$& 2.5 & 3.2& 3.5&1.17\\
$M_{gas}$& 0.5 & 0.9&1.6&0.3\\
$M_{bulge}$        & 2.5 & 1.4&0.4&0.47\\
$M_{DM}$ &128.6&128.6&128.6&42.9\\
\hline
B/T                & 0.5  &0.3&0.1&0.3333              \\
$f_{gas}$          & 0.17 &0.22&0.31&0.22      \\
\hline
$r_{disc}$ & 3.75& 3.79&3.85& 2.63 \\
$z_{disc}$ &  0.75& 0.76&0.77&  0.53\\
$r_{bulge}$ &  0.75& 0.76&0.77&  0.53 \\
$c$ &\multicolumn{4}{c}{9}\\
$\lambda$ & \multicolumn{4}{c}{0.33}\\
\hline
$N_{total}$ &800000&800000&800000&266666\\
$N_{DM}$ &400000&400000&400000&133333\\
$N_{disc, *}$ & 181818&233333&257143&77778\\
$N_{gas}$ & 36364&66667&114286&22222\\
$N_{bulge}$ &       181818&100000&28571&33333\\
\hline
$m_{DM}$ &\multicolumn{4}{c}{$3.2 \times 10 ^6 M_\odot$}\\
$m_{baryon}$ &\multicolumn{4}{c}{$ 1.4 \times 10 ^5 M_\odot$}\\
$\epsilon_{DM}$ &\multicolumn{4}{c}{137\,pc}\\
$\epsilon_{b}$ &\multicolumn{4}{c}{28\,pc}\\
\hline
SFR & $\sim$1 & $\sim$2 & $\sim$5 & $\sim$0.5 \\
\hline
\end{tabular}
\end{table}

The progenitor galaxies are set up following the method given in \citet{Johansson2009equal} assuming a $\Lambda$CDM cosmology with $\Omega_m=0.30$, $\Omega_\Lambda=0.70$, and $H_0=71\,{\rm km/s/Mpc}$, with the aim to mimic disc galaxies observed in the local Universe. A summary of the progenitor galaxy parameters used in this paper can be found in Table \ref{tab:ic_parameters}. The primary galaxies have a virial velocity of $v_{vir}=160\,{\rm km/s}$, resulting in a total virial mass of $M_{vir}=v_{vir}^3/10GH_0=1.34\times10^{12}\ M_\odot$. The majority of this mass  ($M_{DM} = 1.286 \times 10^{12} M_\odot$ ) is in the dark matter  (DM) halo, which has a Hernquist density profile with a concentration parameter of $ c = 9 $  \citep{hernquist1990analytical}. The baryonic mass fraction is 0.041, distributed between a gaseous disc  ($M_{gas}$), a stellar disc  ($M_{disc,*}$) and a stellar bulge  ($M_{bulge}$), with $f_{gas} = M_{gas}/(M_{gas}+M_{disc,*})$ determining the gas fraction in the disc. 
Each progenitor galaxy has a total particle number of $8\times 10^5$. Half are DM particles and half baryonic, yielding a mass resolution of $1.4\times10^5$M$_\odot$ and $3.2\times10^6$M$_\odot$ for baryonic and DM particles respectively. The gravitational softening lengths are $\epsilon_{\rm bar}=28\,{\rm pc}$ and $\epsilon_{\rm DM}=137\,{\rm pc}$ for the baryonic and dark matter components respectively. 

The disc has an exponential mass profile with scale length $r_{disc}$, determined by assuming that the disc material conserves specific angular momentum during the disc formation with a constant halo spin of $\lambda=0.033$ \citep{mo1998formation}. The scale height of the stellar disc is set to $z_{disc}=0.2\,r_{disc}$. The gaseous disc has the same scale length as that of the stellar disc and the vertical structure of the gaseous disc is  determined such that it is in hydrostatic equilibrium  \citep{Springel2005BHmodel}. The stellar bulges follow a \citet{hernquist1990analytical} profile with a scale length of $r_{bulge}=0.2r_{disc}$. The mass of the bulges are determined by the parameter $B/T$, the stellar bulge-to-total stellar mass ratio. By choosing different bulge-to-total stellar mass ratios and corresponding gas fractions, we create 3 progenitor galaxies of different morphology types, aiming to roughly mimic the average properties of Sa, Sc, and Sd galaxies. The specific values are chosen loosely based on values for local SDSS galaxies, by combining the range of $i$-band luminosity B/T values from \citet{Gadotti2009}, with the range of atomic and molecular gas mass fractions from \citet{Saintonge2016}. 

For the simulations with BH feedback the progenitor galaxies host BHs at their potential minima. The mass of the BH is given by the $M_{BH}-\sigma$ relation, with $\sigma$ calculated from the distribution of the bulge star velocities \citep{Bosch2016}.


%

In order to build accurate spectral models, as well as include the feedback from old stars, we must initialise the stellar particles with their ages and metallicities. For the stellar particles in the bulge, we adopt an exponentially decaying star formation rate  (SFR) over the time:
\begin{equation}
SFR_{b} (t)=Ce^{- (t-t_0)/\tau},
\end{equation}
where $t_0=10.2$\,Gyr is the start of the simulation, $C$ is a normalisation factor, and $\tau = 1$ Gyr is the timescale of the exponential decline. We assume that the star formation in the bulge started at a cosmic time of $t=0.5$\,Gyr, leading to a negligible SFR compared to the disc by $t_0$.

For the stellar particles in the disc, we follow \citet{lahen2018fate} and adopt a linearly decaying SFR with an initial SFR at $t_0=10.2$\,Gyr similar to that found in local star-forming disc galaxies. We then run the galaxies in isolation and iteratively adjust the initial SFR estimate until the SFR transitions continuously from the initial value to the actual SFR at the start of the simulation (see Table \ref{tab:ic_parameters}).  
By requiring that the integration of the star formation rate history ove time is equal to the total stellar mass at the start of the simulation, the age distribution of the stellar particles can be obtained.

To be fully consistent with the employed sub-resolution models, we must also initialise the metallicity distribution. Here we adopt a uniform, log-linear radially decaying metallicity profile with the Milky Way as a reference  \citep{zaritsky1994h}. Using the observed Oxygen gradient of 0.0585\,dex/kpc, we set up our galaxies with roughly solar total metallicities. When assigning metal elements to individual stellar particles, we assume a scatter of 0.2 dex motivated by the the maximum measurement error in \citet{kilian1994galactic}
and fix an upper limit on the metal mass fraction of 5\% to prevent the stellar particles from becoming unphysically metal heavy. 

Due to element recycling, stars with later cosmic formation times tend to have higher metallicities. To reproduce this phenomenon, we firstly randomly sample the age of individual stellar particles from the age  distributions with an age scatter of $\Delta t = 100$\,Myr. We then sort the stellar particles by their assigned metal mass fraction, and re-assign their ages keeping the same overall age distribution. 

We additionally create a lower mass galaxy  (Scp3), a smaller version of the Sc galaxy with a mass of one third that of the Sc galaxy. This is done by reducing $v_{vir}$ to $160/\sqrt[3]{3}\,{\rm km/s}\sim111\,{\rm km/s}$. The particle numbers of each component are reduced by one third to keep the mass resolution per particle constant.

\subsection{Merger Simulations}
\label{subsec:merger}

After initialising the progenitor galaxies, we run the galaxy models in isolation for 0.5\,Gyr to even out any numerical artefacts from the initial idealised setup. We then set up the merger simulation with these relaxed galaxies at $t_{m0} = 10.7$\,Gyr, on parabolic trajectories with three different initial orbital configurations as introduced by \citet{naab2003statistical}:
\renewcommand{\labelenumii}{\Roman{enumii}}
\begin{enumerate}
\item G00: $i_1=0^{\circ}$, $i_2=0^{\circ}$, $\omega_1=0^{\circ}$, $\omega_2=0^{\circ}$;
\item G07: $i_1=-109^{\circ}$, $i_2=71^{\circ}$, $\omega_1=-60^{\circ}$, $\omega_2=-30^{\circ}$;
\item G13: $i_1=-109^{\circ}$, $i_2=180^{\circ}$, $\omega_1=60^{\circ}$, $\omega_2=0^{\circ}$.
\end{enumerate}
where $i_1$ and $i_2$ denote the inclinations of the two discs relative to the orbital plane, and 
$\omega_1$ and $\omega_2$ denote the arguments of the orbits' pericenter. G00 is a symmetric prograde-prograde orbit. Both galaxies are orientated in the orbit plane and their angular momenta are parallel to the orbital angular momentum. G07 is a retrograde-prograde orbit with both galaxies inclined with respect to the orbital plane. G13 is retrograde-retrograde orbit, with one galaxy inclined while the other is not. The angular momentum of the second galaxy is anti-parallel to the orbital angular momentum.

The initial separation $r_{\rm sep}$ of the two progenitor galaxies is given by their average virial radii while the pericentre distance $r_p$ is given by the sum of their disc scale lengths. Thus, for the equal-mass mergers  (1:1) the initial separation is $r_{\rm sep} =  (160+160)/2\,{\rm kpc/h} \sim 225\,{\rm kpc}$, the pericentre distance is $r_p=2\times2.7\,{\rm kpc/h} \sim 7.6\,{\rm kpc}$. For the unequal-mass merger  (3:1), the initial separation is $r_{\rm sep} =  (160+111)/2\,{\rm kpc/h} \sim 135.5\,{\rm kpc}$, and the pericentre distance is $r_p=  (2.7+1.90)\,{\rm kpc/h}\sim6.5\,{\rm kpc}$. The progenitor galaxies approach each other following nearly parabolic orbits and interact under their own gravity.

We run a set of simulations with different progenitor galaxies and orbits. For the equal-mass mergers we select 6 illustrative combinations of progenitor galaxies (2xSa, Sa\_Sc, Sa\_Sd, 2xSc, Sc\_Sd, 2xSd), and for the 3:1 mass ratio mergers we choose one high mass progenitor of each type and merge it with the lower mass Sc galaxy giving three further combinations (Sa\_Scp3, Sc\_Scp3, Sd\_Scp3). Each pair is set on 3 different orbits G00, G07, and G13 (Section~\ref{subsec:merger}), giving 18 equal-mass and 9 unequal-mass merger simulations in total.
The exhaustive table of the parameters for the merger simulations can be found in the Appendix (Table \ref{tab:merger_info}).
 
Each merger simulation is run for 3\,Gyr until the current cosmic time $t=13.7$\,Gyr.
Snapshots are output every $2\times10^7$ years, giving 150 snapshots per simulation. 

\section{Building mock MaNGA datacubes}
\label{sec:sedmorph}
We develop the SEDmorph code\footnote{The project web page:  \url{https://github.com/SEDMORPH}. The code for mock datacube creation \url{https://github.com/SEDMORPH/YZCube}.}
standing for \textit{Spectral Energy Distribution and morphology}, to turn the particle data into mock images and integral field datacubes that closely match the properties of the MaNGA survey. We place the mock galaxies at a redshift of $z\sim0.04$, close to the median redshift of the MaNGA survey. The 1\arcsec\ radius MaNGA fibres are equivalent to 0.79\,kpc at this redshift. Here we provide details on the components of the code that are used in this paper. 

\subsection{Stellar continuum creation}
\label{subsec:spec}

To build the stellar continuum spectra we use the \citet[][BC03 hereafter]{bruzual2003stellar} spectral synthesis models, updated to 2016\footnote{These are available at \url{http://www.bruzual.org/~gbruzual/bc03/Updated_version_2016/}}. These models provide integrated light spectra for simple stellar populations  (SSPs) which represent coeval, single metallicity stellar populations, assuming an initial mass function  (IMF), evolutionary tracks and stellar input spectra. This version of the BC03 models are built from both observed and theoretical stellar spectra, assuming a \citet{Chabrier2003} IMF and ``Padova-1994'' evolutionary tracks \citep{Alongi.etal.1993a,Bressan.etal.1993a,Fagotto.etal.1994a,Fagotto.etal.1994b,Girardi.etal.1996a}. The observed spectra most relevant for the wavelength range of the MaNGA spectra are taken from the MILES stellar library \citep{SanchezBlazquez.etal.2006a,FalconBarroso.etal.2011a} in the wavelength range 3540\AA - 7350\AA, extended with the STELIB stellar library \citep{LeBorgne.etal.2003a} out to 8750\AA. 

We assume that each stellar particle in our simulation is composed of a single coeval, fixed metallicity stellar population, with an age and metallicity corresponding to when the progenitor gas particle becomes a stellar particle. The SSPs have already been interpolated onto an optimum age grid, we therefore refrain from any further interpolation in age and each stellar particle is assigned the closest SSP in age. The BC03 models provide SSPs for seven different metallicity values  (with metal mass fractions of $Z=$0.0001, 0.0004, 0.004, 0.008, 0.02, 0.05, 0.1, where $Z_\odot=0.02$). The metallicity distribution of particles in the simulations is such that 
most particles have  metallicities in the 2 bins around solar metallicity ($0.08<Z<0.05$),
which is not sufficient to accurately reproduce the changes in spectral indices caused by the metallicity gradient of the galaxies. We therefore interpolate the spectra linearly in log flux between the central metallicity bins, aiming to distribute the stellar particles between metallicity bins as equally as possible. This leads to additional SSPs with $Z=$ 0.012, 0.016, 0.023, 0.026, 0.035 and 0.04. We verify the success of the interpolation by computing the \hda\ and \hga\ spectral indices of the 1\,Gyr old interpolated spectrum, finding that the indices vary smoothly between the new SSPs as expected.

The stellar particles contain 11 different elements as described in Section~\ref{subsec:subgrid}. We calculate the metallicity of each stellar particle as $Z_* =  (1- M_H - M_{He})/M_*$, where $M_H$ and $M_{He}$ are the total masses of Hydrogen and Helium respectively and $M_*$ is the mass of the star particle, and assign the closest interpolated SSP in metallicity. Finally, the integrated spectrum (luminosity density) of the simulated galaxy or part thereof, can be written as:
\begin{equation}
\label{eq:SPS}
l_{\lambda} = \sum_{i=1}^{N_{*}} M_i \times l_{\lambda,SSP} (t_{SSP, i}, Z_{SSP,i})
\end{equation}
where $M_{i}$ is the mass of the stellar particle, $l_{\lambda,SSP}$ is the luminosity density of the assigned SSP and the sum is over all relevant star particles in the region of interest.  

\subsubsection{Dust attenuation model}
\label{subsec:dust}
We employ a two-component dust attenuation model \citep{charlot2000simple, wild2007bursty} in which the optical depth is contributed by the interstellar medium for all stars, as well as stellar birth clouds for stellar populations younger than $10^7$ years. The final effective optical depths as a function of wavelength for young and old stellar populations are thus given by:
\begin{equation}\label{Equation: dust_model}
  \begin{aligned}
\hat{\tau}_{young} &=& \mu_{d} \tau_{v} \left(\frac{\lambda}{5500}\right)^{-0.7} &+ \left(1-\mu_{d} \right) \tau_{v} \left(\frac{\lambda}{5500}\right)^{-1.3} \\
\hat{\tau}_{old} &=& \mu_{d} \tau_{v} \left(\frac{\lambda}{5500}\right)^{-0.7}  &
 \end{aligned}
\end{equation}
where $\lambda$ is the wavelength in \AA, $\mu_{d}$ is the fraction of optical depth contributed by the ISM, which is set as 0.3 and $\tau_v$ is the effective optical depth at 5500\AA, which we set to a fiducial value of 1.0, typical for local star-forming galaxies.

We note that by not using radiative transfer to calculate the attenuation we are not correctly accounting for the full 3D geometry of the stars and gas in the mock galaxies. Additionally, we do not attempt to link dust content to the metallicity of the gas particles. However, we choose this approach to avoid further sub-resolution recipes given the relatively high mass of the individual gas and star particles compared to giant molecular clouds in which stars are formed in reality. It is important to keep in mind the limitations of any dust modelling in mock galaxies with this level of spatial resolution. 

\subsection{Data Cube Creation}
\label{subsec:datacube}

The first step in creating mock datacubes is to locate the centre of the simulated galaxies in each snapshot. For simulations with black holes we simply take the position of the supermassive black hole as the centre of the galaxy
as the black hole is repositioned to the minimum of the potential at every time step.
For the simulations without black holes we use the distribution of dark matter particles to identify the centre of the potential well in which the galaxy resides. We create a single datacube for each merger simulation, focussing on the centre of the first galaxy, which is the highest mass galaxy in the case of unequal-mass mergers. This is sufficient for our purposes, as we are predominantly interested in the post-merger remnants. 

The basic steps to create a realistic MaNGA datacube are to:  (1) apply a seeing (atmospheric) point spread function  (PSF);  (2) sum the light that falls down each circular fibre in a single pointing;  (3) apply a dither pattern to build the Row Stacked Spectrum  (RSS) file;  (4) combine the RSS into a regularly gridded datacube.
We also explored other algorithms for creating datacubes and found that they are inferior to the one described above.
One of alternative algorithms is to apply the PSF after the cube is built. This method is computationally less expensive by $\sim$20\% but does not lead to sufficiently accurate results.
Another algorithm is to bin the simulation directly onto a square grid and apply the well characterised combined atmospheric and instrumental PSF after the binning. This process is simpler to be realised but turns out to be more computationally expensive.
More details on the comparison of different algorithms can be found in Appendix \ref{app:cubecreation}.

In order to measure the spectrum received by a single MaNGA fibre, we must first apply a PSF to account for the blurring of the light as it travels through the atmosphere. We assume a two-component Gaussian profile with parameters taken from the median of SDSS DR7 imaging fields. For the purposes of this paper we are particularly interested in the spectra of local galaxies around 4000\AA, therefore we apply the typical $g$-band PSF taken from
the median values for 500 000 
SDSS DR7 fields (specifically, the $\sigma$ widths of the two Gaussians are $0.54$\arcsec\ and $1.21$\arcsec\ respectively, and the peak
amplitude ratio of the two Gaussians is $0.081$). 
We calculate the integrated spectrum that falls within the $j$th fibre by summing over stellar particles in the simulation accounting for their distance from the centre of the fibre, $d$, and the shape of the PSF:
\begin{equation}
    l_{\lambda,j} = \sum_{i=1}^{N_{star}}W_{\rm PSF} (d)_i l_{\lambda,i}
\end{equation}
where $l_{\lambda,i}$ is the attenuated stellar continuum luminosity density assigned to the $i$th stellar particle, as described above, and $W_{\rm PSF} (d)$ is a weight function that accounts for the shape of the PSF integrated over the fibre area  (assuming a MaNGA fibre radius of 1\arcsec). In order to reduce the computing time we set $W_{\rm PSF} (d)=0$ once it drops below a value of $10^{-6}$.

The MaNGA fibres are arranged in hexagonal bundles, and sets of three dithered exposures are taken to fill the gaps between the fibres using a triangular pattern with a side length of 1.44\,\arcsec.
In each MaNGA observation, the number of dither sets depends on the on the time it takes to reach the target depth, leading to the an integer multiple of 3 exposures (3,6,9,12, etc).
For our mock observation, one dither set of three exposures will be sufficient.
To know where to place the circular apertures on the simulated galaxy, we take a representative metrology file for the largest MaNGA IFU bundle of 127 fibres (ma134-56995-2.par\footnote{An slightly older version that's publically accessible can be found here: \url{https://svn.sdss.org/public/repo/manga/mangacore/tags/v1_6_2/metrology/ma134/ma134-56995-1.par}. The older version is off by a slight scaling factor from the one used in this paper.}) which records the positions of all the fibres in one dither. We then shift the fibre positions following the triangular pattern for the fibre positions in the other two dithers.

Once we have calculated the spectra in each individual exposure, we combine them into a 3D datacube using the same flux-conserving algorithm as used in the MaNGA survey \citep[see Section 9.1 of ][]{Law2016}. We do not include noise due to e.g. sky background, therefore we do not calculate error arrays or masks. Neither do we model the line spread function, and the resulting spectra remain at the spectral resolution of the SSPs.

\subsection{Spectral analysis}
\label{subsec: specanalysis}

In order to identify post-starburst galaxies we are looking for galaxies with an excess of A/F stars left over from a recent starburst, however ordinary star-forming galaxies have plenty of A/F stars and therefore some knowledge about the current star formation is required in all but the most extreme cases. The traditional method uses a Balmer absorption line  (typically H$\delta$) which is strong in A and F star spectra, alongside an absence of nebular emission to ensure no ongoing star formation. The disadvantage of this method is that it will not select post-starburst galaxies with low-level residual star formation, or those that contain a narrow line AGN  \citep{yan2006, wild2007bursty},
and if a blue emission line is used then samples can be contaminated by dust obscured star-forming galaxies 
\citep{poggianti2000optical, Wild2020star}.
Alternatively we can use the stellar continuum to provide an estimate of the recent star formation, e.g. using a combination of the 4000\AA\ break strength and H$\delta$ absorption line strength  \citep{kauffmann2003stellar}. In more extreme cases, even photometric data can identify the strong Balmer break exhibited by  post-starburst galaxies 
\citep{Whitaker2012,Wild2014b,Forrest2018}. 
Here we compare two methods for identifying post-starburst regions in our simulated galaxies. 

\subsubsection{The traditional method}
\label{subsec:spec_indices}
We sum the \hda\ and \hga\ absorption line indices \citep{HdA} and combine this with the equivalent width of the H$\alpha$ emission line to investigate when the simulated galaxies obey the ``traditional'' selection criteria for post-starburst galaxies. 
We measure the \hda\ and \hga\ absorption indices using the same method as in the MaNGA survey (see Section \ref{sec:observations}). We note that these are calculated at the resolution of the MILES-SSPs, $\sim$58\,km/s, which is similar to the MaNGA instrumental dispersion around the 4000\,\AA\ break. As we are dealing with small regions of galaxies, the velocity dispersion is not as large as for whole galaxies. Hence, the slight difference in resolution between the simulated spectra and the MaNGA data has little effect on the spectral indices used in this paper.

In order to calculate the equivalent width of the \ha\ emission line consistently from the spectral synthesis models, we firstly measure the ionising photon luminosity $Q (H^0)$ from the dust-free continuum luminosity density $l(\lambda)$ (Eqn.~\ref{eq:SPS}) at wavelengths shorter than the Lyman limit ($\lambda_{ly}$):
\begin{equation}
Q ({\rm H}^0) = \int_{}^{\lambda_{ly}}l(\lambda) \frac{\lambda}{hc} \mathrm{d}\lambda
\end{equation}
where  $h$ is the Planck constant and $c$ is the speed of light. We convert this into an integrated \ha\ emission line luminosity  ($L (H\alpha)$) assuming Case B recombination at $T_e = 10,000K$ \citep[e.g.][]{kennicutt1998star}:
\begin{equation}
	L ({\rm H}\alpha) [\textrm{ergs s}^{-1}]=1.37x10^{-12}\times Q ({\rm H}^0) [\textrm{s}^{-1}].
\end{equation}
We then attenuate the luminosity due to dust by the same amount as for young stars in our continuum model (Eqn. \ref{Equation: dust_model}). To calculate the equivalent width (EW) of the \ha\ emission line as used in observations, we extract the mean stellar continuum luminosity density from the attenuated stellar continuum between $\pm5$\AA\ from the wavelength of \ha\ ($l_{6563}$). The EW is then given by: 
\begin{equation}
    W({\rm H}\alpha) = \frac{L({\rm H}\alpha)}{l_{6563}}.
\end{equation}

\subsubsection{Principal Component Analysis}
\label{subsec:PCA}
We compare the traditional method with a second spectroscopic diagnostic which uses the stellar continuum alone to identify galaxies with stronger Balmer absorption than expected. This method applies a Principal Component Analysis (PCA) to the stellar continuum between 3750 and 4150\AA, finding the first component (PC1) to correlate strongly with the 4000\AA\ break strength, while the second component (PC2) provided the excess Balmer absorption line strength over that expected for the 4000\AA\ break strength  \citep{wild2007bursty}. The method is  analogous to using $D_n4000$ vs. \hda, however has the added benefit of being able to combine information from all the Balmer absorption lines as well as the shape of the continuum in order to increase the SNR of the measurement, as well as rotating the parameter space so that the PSBs are easily identified lying above the star-forming main sequence. This means that older and weaker PSBs can be detected with the PCA compared to \hda\ alone \citep[see ][ for a direct comparison]{wild2007bursty}. As the emission lines are not used, the method is sensitive to galaxies that contain narrow line AGN, and those galaxies that do not completely shut off their star formation. 

In order to calculate the principal component amplitudes for the spectra in the mock data cubes, we first convert from air to vacuum wavelengths, convolve to the common velocity dispersion of 150\,km/s used by the eigenbasis (assuming an intrinsic dispersion of 58\,km/s for the MILES library) and then project the spectra onto the same eigenvectors as calculated in \citet{wild2007bursty}\footnote{The eigenvectors can be downloaded here: \url{http://www-star.st-and.ac.uk/~vw8/downloads/DR7PCA.html}. Code to perform the projection is available here: \url{https://github.com/SEDMORPH/VWPCA} (IDL) or \url{https://github.com/astroweaver/pygappy} (python).}

\section{Results}
\label{sec:results}

In this section we compare the impact of different black hole feedback models, progenitor galaxy properties and orbits on the star formation histories and spectral measurements of the simulated galaxies as a function of time from pre- to post-merger. We then focus in on one particular simulation to study the origin of the radial gradients in the spectral indices. 

\subsection{Comparison between different BH feedback models}
\label{subsec:compare_BH}
\begin{figure}
	\includegraphics[width=0.95\columnwidth]{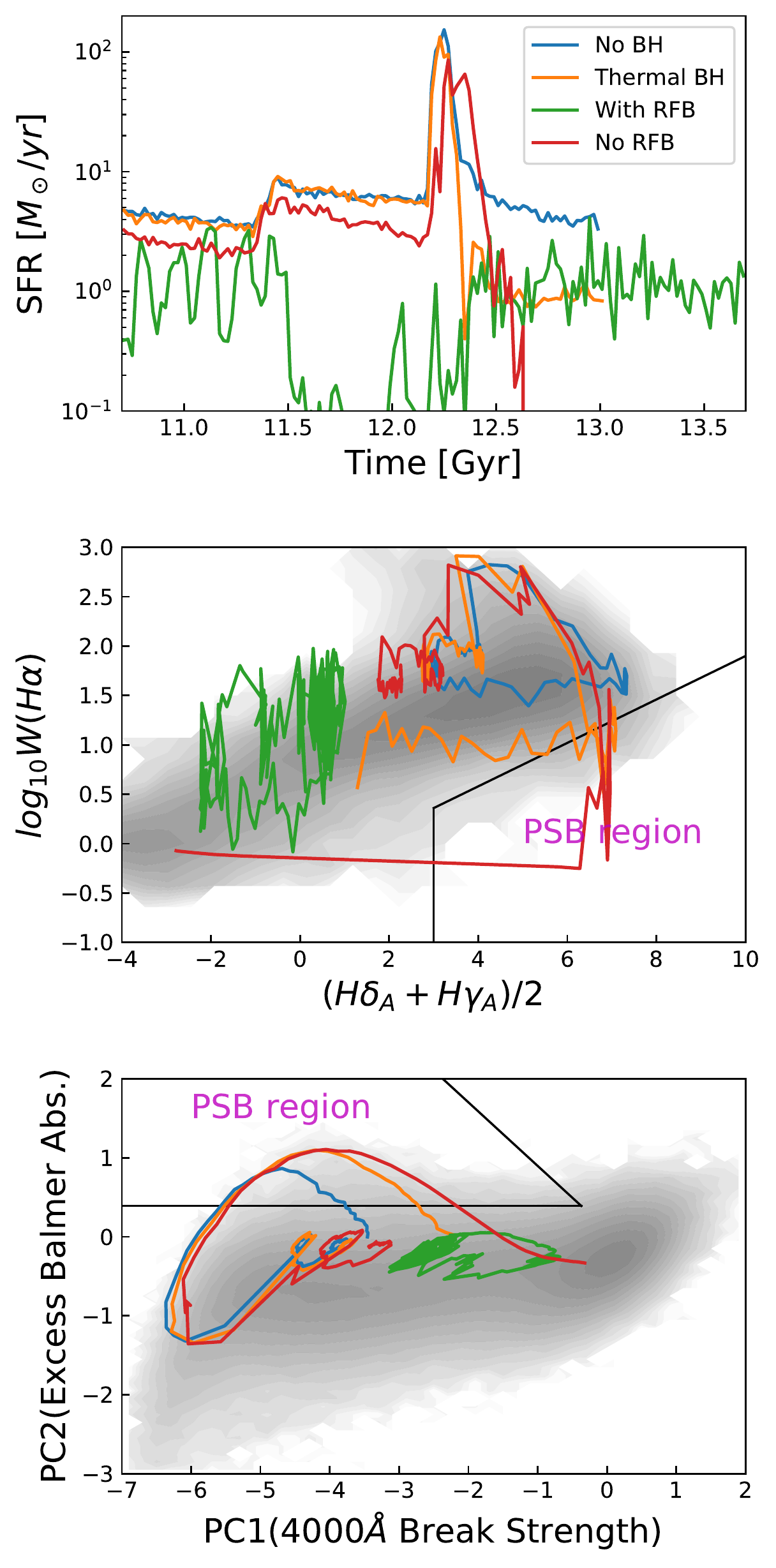}
	\caption{Comparison of simulations with the same progenitor galaxies and orbits (2xSc\_00), but different BH feedback models. In all panels  the different colour lines show a model with: the BH feedback turned off (blue, ``\textit{No BH''}); ``classical'' thermal BH feedback model (orange, ``\textit{Thermal BH}''); mechanical BH feedback with additional X-ray radiative feedback (green, ``\textit{With RFB}''; mechanical BH feedback without additional X-ray radiative feedback (red, ``\textit{No RFB}''). In the \textit{No BH} and \textit{Thermal BH} simulations the model galaxies run out of gas particles before the end of the simulations, thus entering a numerically induced quenching phase. We therefore only show the evolution of these simulations up to that point.
	\textit{Top}: The star formation histories. \textit{Middle}: The evolution of the simulated galaxies in the Balmer absorption and \ha\ emission line space. \textit{Bottom}: The evolution of the simulated galaxies in PC space. In the central and lower panels the underlying grey contours show the distribution of the SDSS DR7 galaxies using a log scaled number density and the black lines indicate the commonly used demarcations for identifying clean samples of PSB galaxies. 
	}
	\label{fig:compare_models}
\end{figure}
In this subsection we compare the quenching progress and observational signatures of quenching in simulations with very different BH models in order to constrain the type of feedback that might be needed to create the post-starburst galaxies observed in the real Universe. Here we consider four models:
\begin{enumerate}
	\item \textit{No BH}: the BH feedback is turned off; 
	\item \textit{Thermal BH}: the ``classical'' thermal BH feedback model (Section~\ref{subsec:oldBH});
	\item \textit{With RFB}: the mechanical BH feedback model with additional X-ray radiative feedback turned \textbf{on} (Section~\ref{subsec:newBH});
	\item \textit{No RFB}: the mechanical BH feedback model with additional X-ray radiative feedback turned \textbf{off}.
\end{enumerate}
We compare simulations with the same progenitor galaxies (2xSc) and orbits (G00); this orbital arrangement produces the strongest tidal forces and therefore cleanest star formation histories making them ideal to compare the impact of the different feedback mechanisms. The star formation histories (SFH) are derived from the change in the total stellar mass of the galaxies between different snapshots and the spectral indices are measured as described in Section~\ref{subsec: specanalysis}. 

The results are shown in \autoref{fig:compare_models}. All simulations are run for 3\,Gyr,
however some simulations experience a very sudden quenching phase caused 
by a complete exhaustion of cold star-forming gas. 
This is likely caused by a combination of the employed supernova$+$BH feedback model and the relatively low particle resolution of our simulations.
We argue that such "numerical induced quenching" is not astrophysical.
To avoid confusion we remove these results from the figures. 
In all cases apart from the model with both mechanical and radiative feedback (\textit{With RFB}) the SFH is dominated by a strong burst of star formation at coalescence ($t\sim$12.3\,Gyr), which leads to post-starburst features in one or both of the observational index spaces. However, the models differ in the amount that they quench and therefore the final SFR as we will discuss below. 
 
It is clear that the mechanical BH feedback model with radiative feedback turned \textbf{on} is too effective at suppressing the star formation. During the first encounter of the galaxies at $t\sim11.5$\,Gyr some gas is funnelled into the galaxy centre and feeds the BH. The radiative feedback strongly suppresses the star formation below 0.1\,\sfrunit, which prevents the small increase in SFR seen in each of the other simulations. Moreover, the radiative feedback completely eliminates the starburst during the final coalescence, preventing the remnant galaxy from showing observational post-starburst features.  After the final merger, the galaxy structure becomes relative stable and the gas inflow becomes weaker, which limits the strength of the BH feedback. In observational index space, the galaxy sits predominantly in the ``green-valley'' throughout the simulation, temporarily reaching the red sequence after the first encounter. The majority of the gas in the galaxy remains unconsumed by the end of the simulation allowing a relative high final SFR of about 1\,\sfrunit, and it quickly returns to the ``green valley'' following coalescence. 

In the other three simulations, the SFHs show a similar trend before the final merger. All the galaxies experience enhanced star formation during the first encounter of the two progenitor galaxies and a strong starburst during the final merger. 
Note that although the initial SFRs are set to be the same when creating the model galaxies,
the SFRs end up with slightly different values when the galaxies reach equilibrium states after the 0.5\,Gyr isolated run (see Section \ref{subsec:merger}).
Shortly after coalescence the SFRs decline quickly from the peak, even in the case of no BH feedback. As noted previously in the literature, this initial decline is caused by gas consumption and stellar feedback from O and B-type stars, rather than BH feedback processes. All three galaxies enter the fiducial PSB region in PCA space, due to the spectral signatures of the excess of A/F stars created in the starburst. Following coalescence we see the divergence in the SFH caused by the different BH feedback models. Without BH feedback (\textit{NoBH}) the remnant's SFR  remains at a level comparable to that before the merger. Introducing thermal feedback suppresses star formation slightly in comparison, but only to the level of 1\,\sfrunit\ (i.e. similar to the Milky Way). Only the mecahnical BH feedback model \textit{No RFB} is able to completely quench star formation in the galaxy. Due to the residual star formation in the \textit{No BH} and \textit{Thermal BH} models these galaxies retain a level of star formation that leads to significant \ha\ emission and prevents the galaxy from passing into the traditional PSB selection region. Only the mechanical feedback model is able to suppress the \ha\ emission sufficiently that the galaxy would be selected as a PSB using the traditional \ha\ emission vs. Balmer absorption indices. The merger remnant evolves onto the red sequence in both index spaces as the simulation ends. 

Based on the discussion above, for the remainder of this paper we focus on simulations with mechanical BH feedback implemented, but without radiative feedback, as these are the only runs that would be selected using a traditional PSB selection method. We note that the fact that the PCA selection includes objects where the quenching is incomplete may well explain the much larger number of PCA-selected PSBs compared to those selected with the traditional cut on emission line luminosities \citep{pawlik2018origins}. 

\subsection{Comparison between different progenitor galaxies and orbits}
\label{subsec:all_SFH}
\begin{figure*}
	\includegraphics[width=1.8\columnwidth]{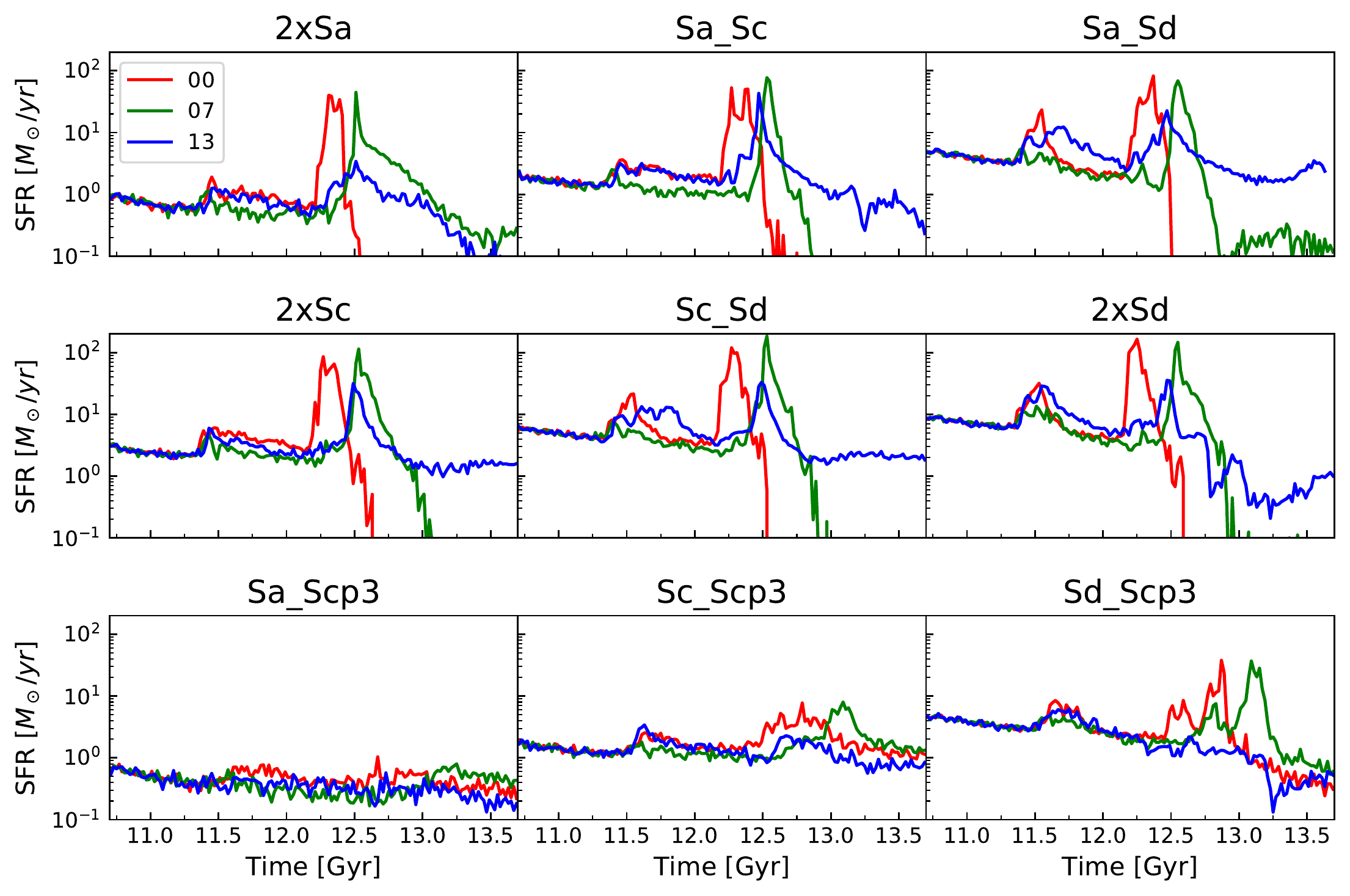}
	\caption{SFH of all the merger simulations run with the mechanical BH feedback model and without the additional X-ray radiative feedback. The star formation histories of the merger simulations demonstrate that sharp quenching is only achieved in particular circumstances: progenitor galaxies with similar mass, approaching each other in either prograde-prograde or retrograde-prograde orbits. Neither unequal-mass merger nor retrograde-retrograde merger can reproduce fast quenching.
	}
	\label{fig:all_SFH}
\end{figure*}

We now run the complete set of simulations presented in Table \ref{tab:merger_info} with different progenitor galaxies and orbits, using the mechanical BH feedback model. 
There are 6 combinations of progenitor galaxies for equal-mass merger  (2xSa, Sa\_Sc, Sa\_Sd, 2xSc, Sc\_Sd, 2xSd) and 3 combinations for 1:3 mass ratio mergers (Sa\_Scp3, Sc\_Scp3, Sd\_Scp3).
Each pair is set on 3 different orbits G00, G07, and G13 (Section~\ref{subsec:merger}
giving 27 different mergers in total. 
In \autoref{fig:all_SFH} we show the SFH of all 27 simulations. We refrain from showing all evolutionary plots of the spectral indices, as the results can largely be inferred from the SFHs given the knowledge gained in the previous subsection. In this subsection we summarise the results for all 27 simulations, and in the following subsection we focus on one key example showing the full evolution of the model in the observational parameter space. 

Almost all equal-mass mergers have strong starbursts, and the higher the gas fraction the stronger the burst. However, despite the strong mechanical BH feedback the starburst can have a long decay time following the merger, especially for the retrograde-retrograde (G13) merger or the retrograde-prograde (G07) merger with lower gas fractions. In these cases, the SFR displays only a gradual decline, and the \ha\ EW never declines sufficiently to place the galaxy into the PSB selection box. For unequal-mass mergers, the starburst is significantly weaker compared to the equal-mass merger with the same orbits and progenitors with the same Hubble type. A higher gas fraction again leads to a stronger starburst, however only the Sd\_Scp3\_00  and Sd\_Scp3\_07 mergers enter the PCA PSB region, although with a lower PC2 than the major mergers. 
Moreover, in all of the nine unequal-mass merger simulations, the post-merger SFR is comparable to the SFR before the merger, leading to strong residual \ha\ emission. 

The prograde-prograde orbit  (G00) is the easiest orbit to reproduce a strong starburst and subsequent quenching. In this orbit, the stellar disc and bulge of the two progenitors collide with each other violently, causing the strongest tidal forces which drive a rapid flow of gas into the galaxy centre forming a large number of new stars in a short time. The gas is quickly consumed, leaving less material for further star formation. The gas flow drives rapid AGN accretion and the resulting BH feedback also reduces the efficiency for the remaining gas to transform into stars. A sharp decline in SFR can be seen in every pair of the equal-mass prograde-prograde progenitors. However, it seems unlikely that this orbital configuration occurs regularly in the real Universe. Comparison with very large cosmological simulations would be required to clarify this. 

We conclude that, within the scope of these simulations, sharp and sustained quenching of star formation caused by mergers is only achieved in particular circumstances: relatively gas rich progenitor galaxies with similar mass, approaching each other in either prograde-prograde or retrograde-prograde orbits. In these simulations, neither unequal-mass nor retrograde-retrograde mergers can reproduce features completely consistent with the full range of observed PSB galaxies, with the presence of nebular emission due to ongoing residual star formation being a key constraint. 

\subsection{Evolution of global spectral properties for a representative merger}
\label{subsec:gal_property}

\begin{figure}
	\includegraphics[width=\columnwidth]{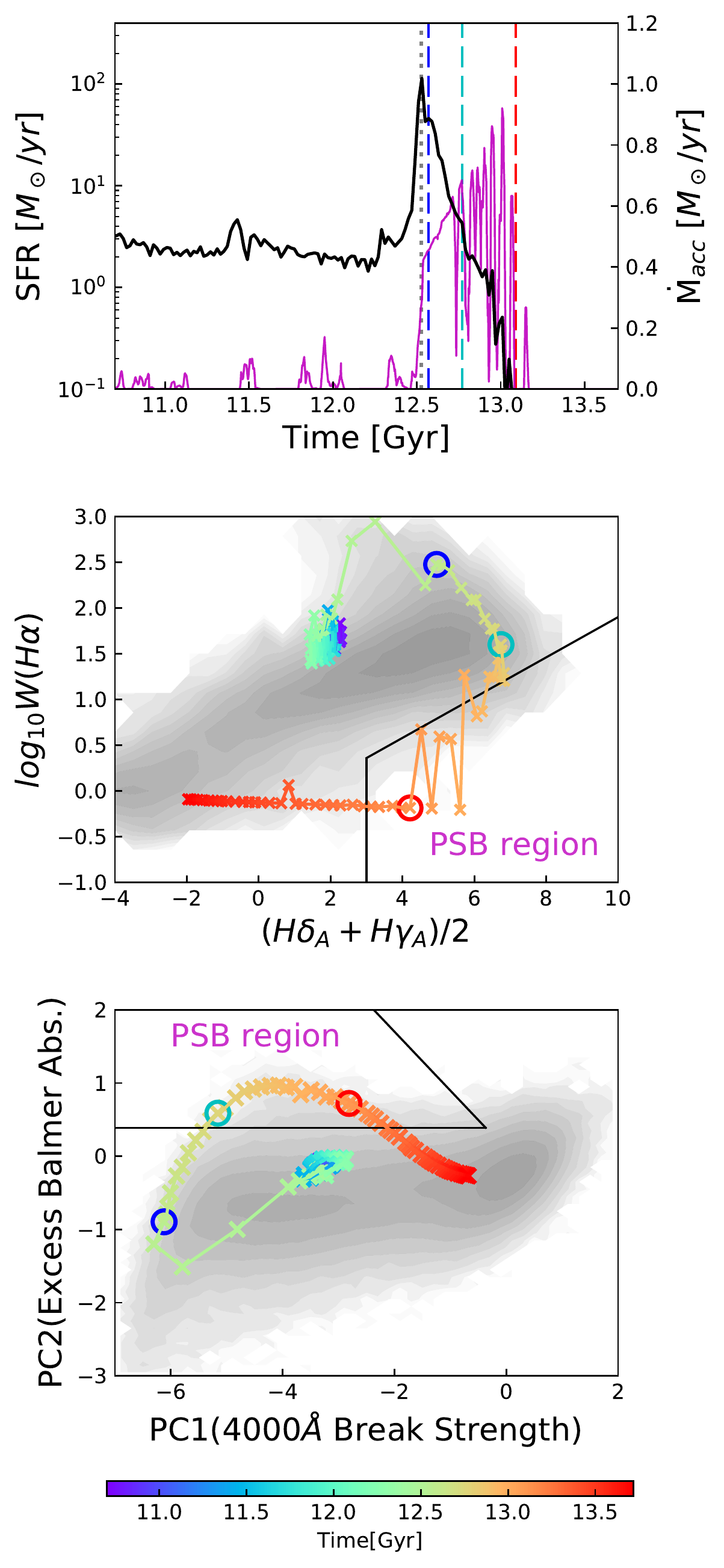}
    \caption{The evolution of the global SFH and spectral properties in the \themerger\ simulation.  The blue, cyan and red dashed vertical lines and circles indicate the snapshots of different quenching stages ($t=12.56$, 12.76, and 13.08\,Gyr respectively). The grey dotted line indicates the time of BH coalescence. 
   	\textit{Top:} the star formation history (black) and BH accretion rate (purple). \textit{Centre:} the evolution of the simulated galaxy spectrum in Balmer absorption vs. \ha\ emission strength. Each cross represents a simulation snapshot spaced every $2\times10^7$ years, with colour from blue to red indicating the progression of time over 3\,Gyr with the colour scale given at the bottom of the figure. \textit{Bottom:} the evolution of the simulated galaxy spectrum in the PCA space. }
    \label{fig:gal_pro}
\end{figure}

The previous two subsections have shown that post-starburst features can be reproduced with specific progenitor galaxies and certain orbits, in the presence of significant BH feedback. We now choose one representative merger simulation and focus on its evolving spectral properties. Though the prograde-prograde orbit (G00) produces the strongest starburst and sharpest quenching, such an orbital configuration with both progenitor galaxies having angular momenta parallel to the orbital angular momentum is likely to be rare in the local Universe. We therefore select the next best G07 orbit. Similarly, mergers between two massive disc-dominated galaxies with high gas fractions are unlikely to be common enough to be the most representative cases. We therefore select the Sc progenitors. In this and the following subsections we thus focus on the \themerger simulation.

The SFH and BH accretion rate of the combined simulation cube are plotted in the top panel of \autoref{fig:gal_pro}, with the three coloured vertical lines indicating the different quenching stages. At the blue snapshot ($t=12.56$\,Gyr), the SFR starts to decrease  (the quenching just begins); at the cyan snapshot ($t=12.76$\,Gyr), the SFR drops to a value that is comparable to that before the merger; at the red snapshot ($t=13$\,Gyr), the SFR drops below 0.1\sfrunit. The dotted grey line indicates the merger of two BHs at $t=12.53$\,Gyr, which is considered as the formation point of the merger remnant. 

The central and lower panels of \autoref{fig:gal_pro} show the evolution of the simulated galaxies in spectral index space, integrating the entire spectrum of both galaxies. The simulated galaxy starts from within the blue cloud in both index spaces at the beginning of the simulation, with the first encounter causing a very small change in the spectral indices. During the starburst we observe very strong \ha\ emission and weak 4000\AA\ break strength (PC1) due to significant contribution to the spectrum from O and B-type stars. The Balmer absorption lines remain strong in starburst galaxies, due to the large number of A and F-type stars formed which are only partially outshone by the O and B stars. However, this translates into a low PC2 as the Balmer absorption line strength is slightly weaker than expected when compared to normal star-forming galaxies. As the star formation declines below its initial level, the remnant quickly moves into the PSB region in both spectral index spaces, ending up on the red sequence. The remnant galaxy can be found in the PCA-defined PSB region consistently from from 12.74 to 13.22\,Gyr, while it is found less consistently in the \ha-defined PSB region between 12.92 and 13.16\,Gyr.

While the stellar continuum based PCA indices show a stable evolution with time, the \ha\ emission line strength fluctuates significantly during the PSB phase. The purple line in the top panel of \autoref{fig:gal_pro} shows the smoothed BHAR averaged
over 20\,Myr,
which fluctuates substantially with peaks in BHAR followed by dips in the SFR. 
We therefore suggest that the fluctuating BH feedback strength causes the galaxy to shift in and out of the \ha-defined PSB region even after entering the PSB phase at $t\sim 12.92$\,Gyr. 

In general for our simulations, we find that the stellar continuum based PCA selection is a more consistent identifier of PSB galaxies. The very strong fluctuations in the \ha\ EW will cause an emission line-selected sample to be incomplete, even before the loss of objects from the sample due to narrow line emission from the AGN. However, we note that some observed PSB galaxies selected using the traditional nebular emission based method are not selected with the PCA, in particular the two examples presented in \autoref{fig:observations}. We also note that this is not an intrinsic property of the PCA analysis, but caused by not using emission lines in the selection. It would be equally true of any PSB selection method that does not use emission lines. We will return to these points in the discussion section below. 

\subsection{Evolution of radial gradients in spectral properties}
\label{subsec:gradient}

\begin{figure}
\centering
\includegraphics[width=\columnwidth]{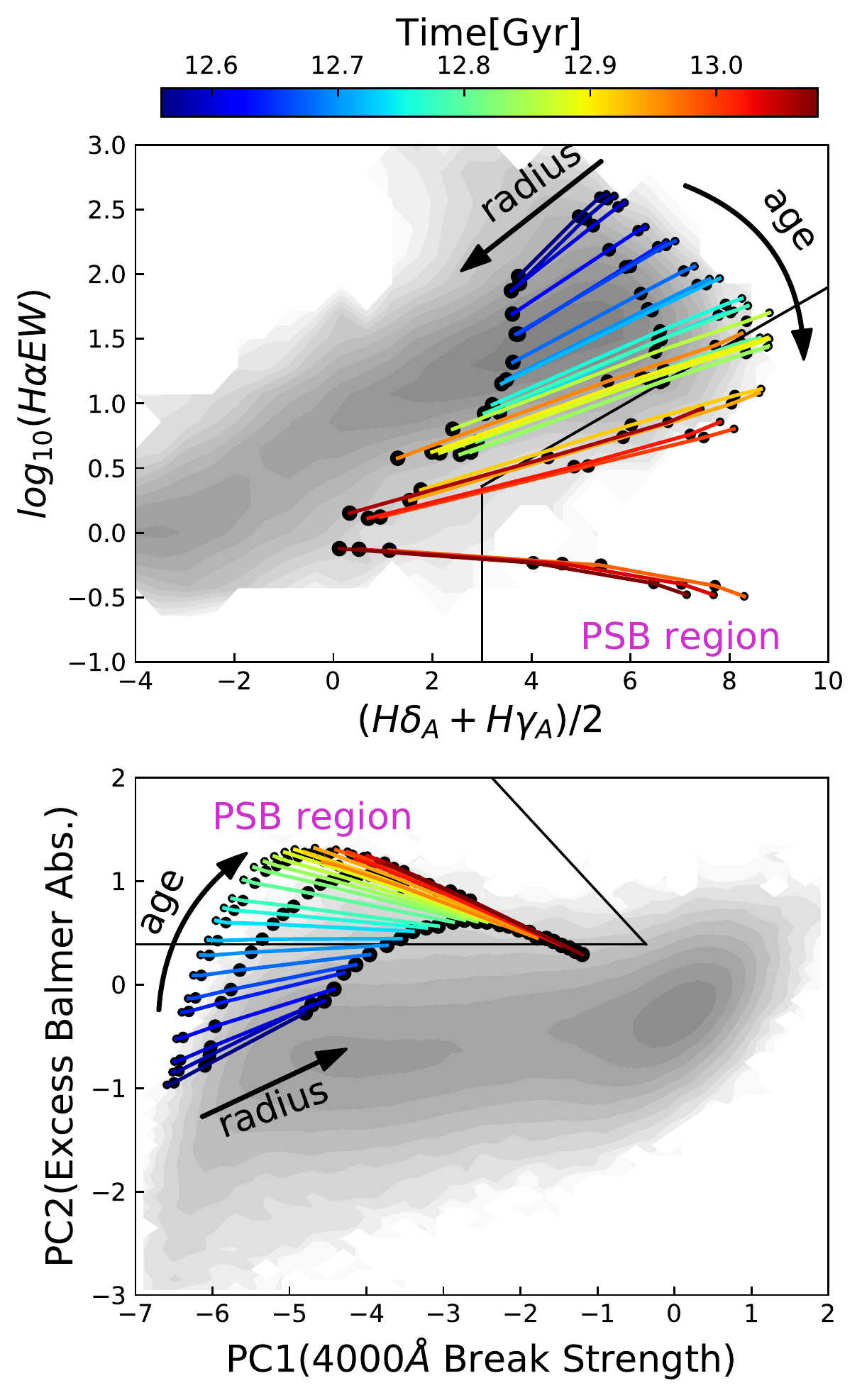}
	\caption{The radial gradients in spectral indices for simulation \themerger between $t= 12.56$ and $13.08$\,Gyr (i.e. the time range between the blue and red lines in the top panel of \autoref{fig:gal_pro}). The spectra are integrated within circular annuli of radius 0--1, 1--2, 2--3, and 3--4\,kpc respectively, shown by small to large dots respectively. The colour of the lines indicates the simulation time as given by the colour bar on the top. Note that in the \themerger simulation, the BHs merge at $t\sim12.53$\,Gyr.
	\textit{Top:} the evolution of the radial gradient in \ha\ emission line equivalent width vs. Balmer absorption line strength. \textit{Bottom:} the evolution of the radial gradient in the PCA stellar continuum indices.  
	}
	\label{fig:gradient}
\end{figure}

\begin{figure}
	\includegraphics[width=\columnwidth]{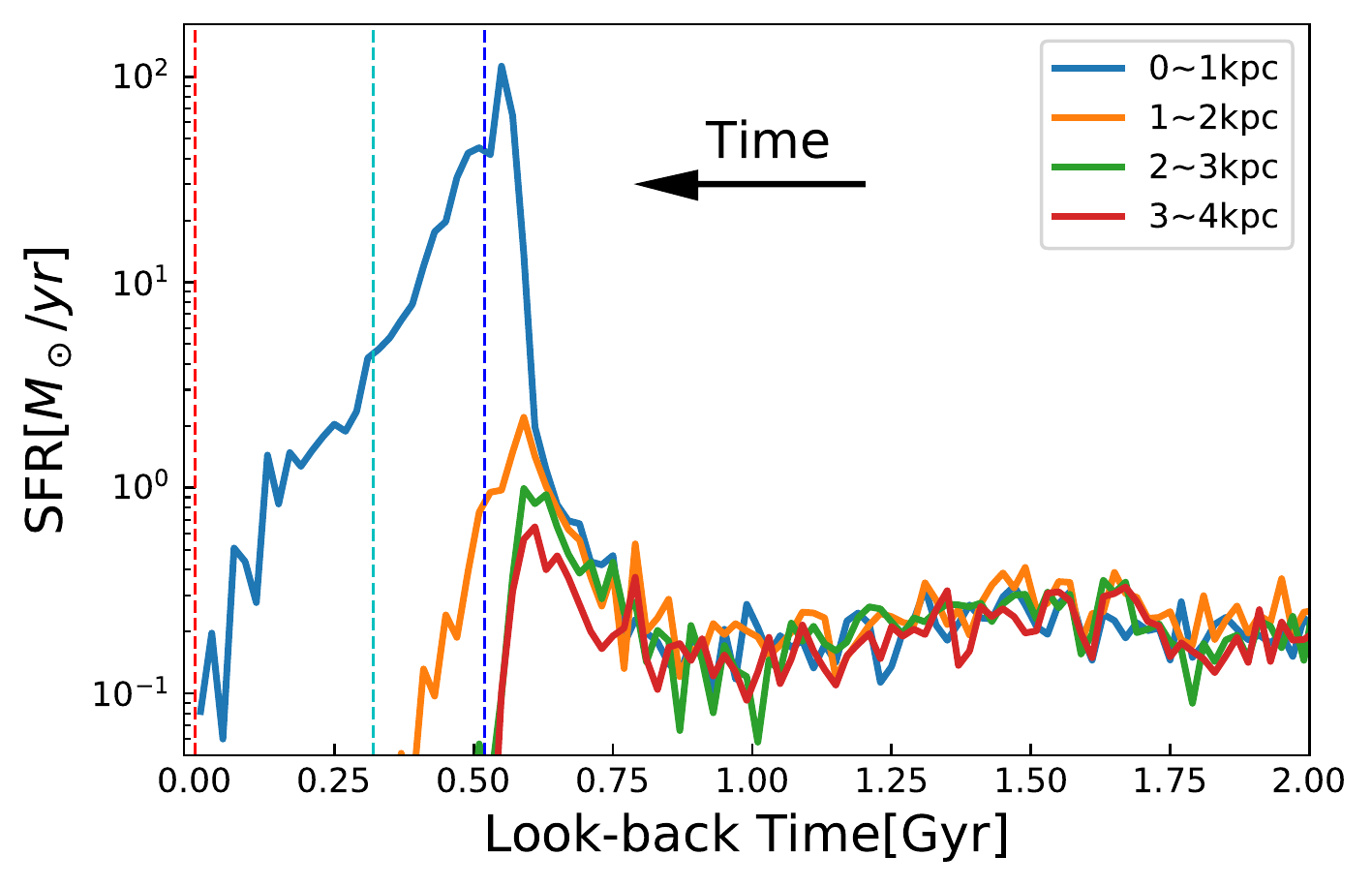}
	\caption{The star formation history in different radial annuli of the galaxy, looking back from $t=13.08$\,Gyr, where the SFR decays below 0.1\sfrunit. The dashed lines are as in \autoref{fig:gal_pro}. The star formation rates peak at roughly the same time at all radii but the burst is significantly stronger and more prolonged in the galaxy centre. 
	}
\label{fig:SFH_radii}
\end{figure}



	
To analyse the spatial properties of the PSB remnant in the \themerger\ simulation, we create datacubes for the snapshots between $t= 12.56$\,Gyr, and $t= 13.08$\,Gyr, from the early quenching stage to the late quenching (i.e. the time range between the blue and red lines in the top panel of \autoref{fig:gal_pro}). To simplify the analysis we bin the spectra in the mock datacubes according to their distance to the galaxy centre, in annuli of 0--1 kpc, 1--2 kpc, 2--3 kpc, and 3--4 kpc respectively, and calculate the spectral indices from the binned spectra. The evolution in the radial gradients of the spectral indices are plotted in \autoref{fig:gradient}. As the \ha\ emission line is very sensitive to the fluctuating residual SFR in the post-merger galaxies, the radial gradient in \ha-\hda\ fluctuates substantially with time during the PSB phase. There is a very strong radial gradient in the Balmer absorption lines at all snapshots, with the central region showing the strongest Balmer absorption lines. In all but a few snapshots, the central region also shows the strongest nebular emission lines. In PCA space the evolution is more stable with time, evolving rapidly from a positive to negative gradient, with the galaxy entering the fiducial selection box at the same time for all radii. In both spectral index figures, only the inner regions of the galaxy are clearly identified as post-starburst, while the outer regions move from star-forming towards quiescent stellar populations. 

The ability of the simulation to reproduce the very different radial gradients observed in different PSB galaxies shown in \autoref{fig:observations} suggests that the difference in observed gradients may simply be due to the time at which we catch the galaxy following coalescence. In Section \ref{sec:observations} we noted that the radial gradients could be produced by a single co-eval burst which was stronger in the central regions, or a starburst that has progressed from outside-in. However, simple toy models are unable to distinguish between these two hypotheses (Weaver et al. in prep.).  In \autoref{fig:SFH_radii} we show the star formation history of the different radial annuli, taken from the final snapshot shown in \autoref{fig:gradient} with $t=13.08$\,Gyr.  The star formation rates peak at roughly the same time at all the radii but the burst is both stronger and more prolonged at the galaxy centre i.e. a combination of the co-eval starburst and the outside-in starburst hypothesese. While the peaks of the star formation are co-eval, the star formation is quenched in the outer regions more quickly than in the inner regions, presumably due to the rapid gas flows continuing to feed the central regions. 

\subsection{Spectral index maps}

\begin{figure*}
    \centering
    \includegraphics[width=\textwidth]{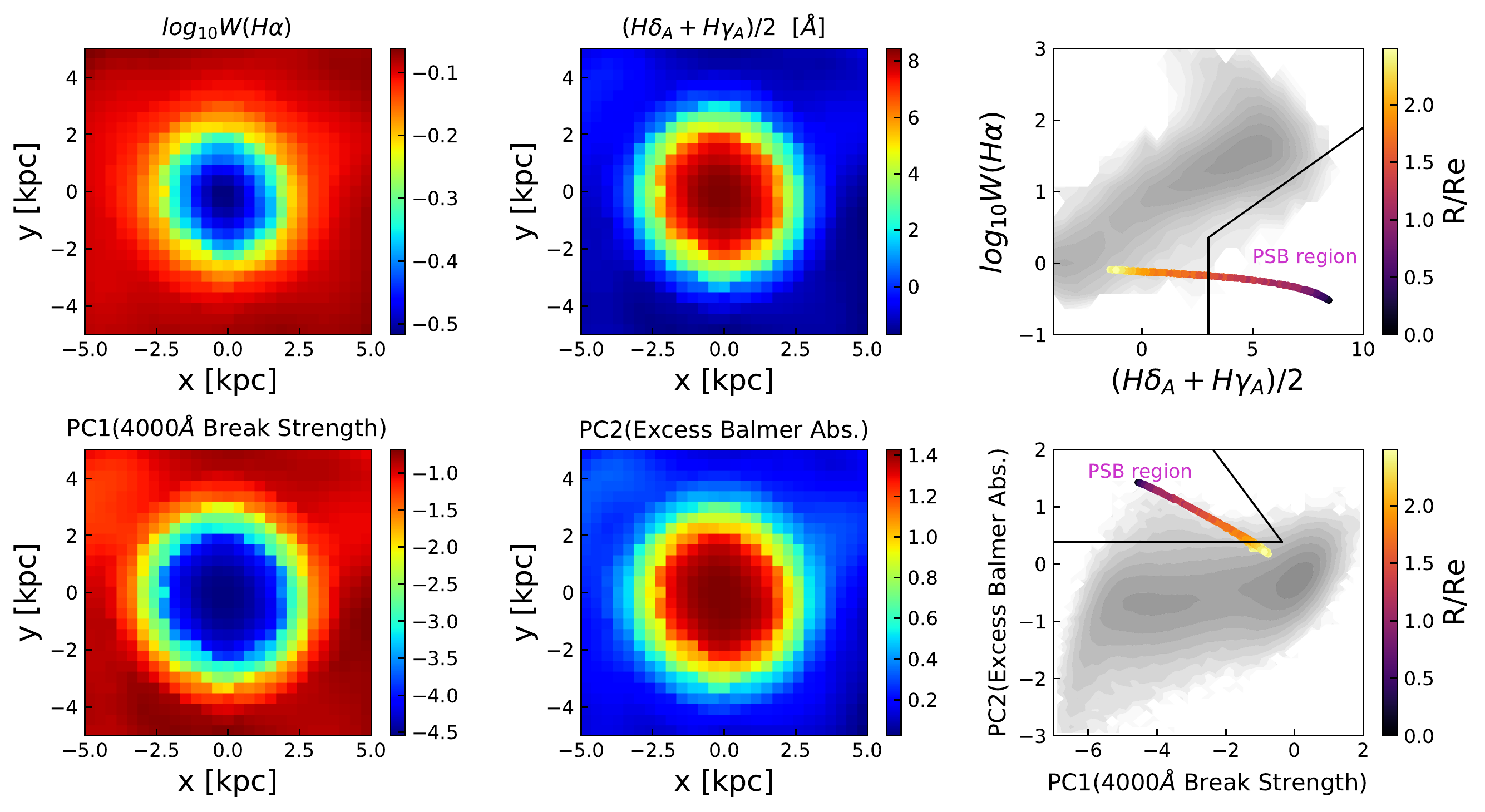}
    \caption{Spatially resolved maps of the key spectral indices used in this paper, for a single snapshot of simulation \themerger chosen for its strong gradient in PC1/2 and minimal ongoing star formation leading to a flat gradient in W(\ha). On the right we reproduce the radially averaged gradients, colour coded by distance from the centre of the galaxy.}
    \label{fig:mockmaps}
\end{figure*}

For completeness, we present the 2D maps of one snapshot of merger \themerger in \autoref{fig:mockmaps}. The snapshot was chosen for the maximal strength of its post-starburst spectral features. We see that the spatial distributions of the spectral indices are entirely smooth, with very little residual structure remaining following the catastrophic event that caused them.

\section{Discussion}
\label{sec:Discussion}

A better model of the physical mechanisms responsible for creating post-starburst galaxies may help to improve our understanding of how and why star formation completely quenches in galaxies throughout cosmic time. Through a careful comparison of real observations from state-of-the-art IFU surveys and mock observations from advanced simulations, we hope to be able to constrain the external and internal processes responsible for quenching star formation. 

The galaxy merger simulations presented in this paper were designed to reproduce low redshift central post-starburst galaxies, where there is significant evidence that major mergers are the only plausible mechanism for the creation of the strongest spectral features and the high fraction of morphological disturbance (see Section \ref{sec:intro}). However we have shown that with the galaxy merger simulations presented here it is difficult to create a sharp enough quenching event that leads to both the strong Balmer absorption lines and the complete absence of nebular emission lines due to ongoing star formation. While the strong gas inflow is required to drive the starburst, it is hard to completely shut off the star formation in the centre of the merger remnant without very significant BH feedback, for which there is limited evidence for galaxies in the very local Universe \citep{fabian2012}.
Whether this is due to the limited size of our simulation set, or limitations in the sub-resolution physics of the models is unclear. Comparison with a full cosmological hydrodynamical simulation would address the former concern, however the latter is much more difficult to tackle: the simulations presented here use very similar recipes to the EAGLE simulation, and most of the currently available simulations do not differ significantly in any of the key ISM or feedback aspects. 

One obvious concern is the observed presence of cold gas in a large fraction of PSBs which appears to have very low star formation efficiencies \citep{rowlands2015, french2015, smercina2018, Li2019}. For simulations that assume a Kennicutt-Schmidt style star formation law, cold gas forms stars, without any exceptions. In addition, our simulations assume a relatively smooth dust distribution, while it is possible that the observed EW of emission lines is reduced by a highly clumpy dust distribution. Improved recipes for treating the detailed sub-resolution physics of the ISM in these unusual galaxies may thus be required. 

Our results clearly disagree with the conclusions of \citet{snyder2011k+} and \citet{wild2009post}: with the improved simulation resolution and sub-resolution physics recipes we find that stellar feedback is only able to cause the initial decline of SFR needed to halt the starburst. Without BH feedback the galaxies return to the blue-cloud or green valley, but cannot further quench the star formation to become entirely quiescent.  

This leads us to the question of whether rapid quenching is related to mergers at all. While it is hard to envisage an alternative scenario in the local Universe, the higher gas mass fractions of galaxies at higher redshift may provide the fuel needed for galaxy wide starbursts, without the need for strong centralised inflows which are difficult to halt. \citet{rodriguez-montero2019} used the SIMBA cosmological simulations to show that in most cases the quenching of the galaxy was not related to a recent major merger, over a wide range of cosmic time. A key additional constraint not investigated here is the morphologies of the remnants: highly compact at high redshift \citep{Yano2016,Almaini2017} and tidally disturbed at low redshift \citep[e.g.][]{zabludoff1996environment, yang2008, pracy2009}. \citet{pawlik2018origins} carried out a first look at the morphological asymmetries of mock images created from very similar simulations to those used in this paper, finding them to decay very rapidly at SDSS-like image depths. Further progress may be possible with deeper imaging datasets, such as will become available in the near future with the 
Vera C. Rubin Observatory.

With the advent of IFU datasets, the differing radial gradients in the spectral indices of PSB galaxies have become apparent \citep{pracy2005,Chen2019}. A negative gradient in the H$\delta$ absorption line strength (i.e. stronger absorption in the centre) as seen in the vast majority of central PSBs \citep{Chen2019}, has commonly been suggested to be an indication of a merger origin due to the gas inflows leading to an excess of A and F stars in the centre. This has been borne out by merger simulations, including those presented here \citep{Bekki2005,snyder2011k+}. Galaxies with PSB spectral features only in their outskirts (so-called `ring' PSBs) are not produced at all by our suite of simulations, and are more likely caused by external processes which have disrupted the outer disc star formation \citep{Owers2019}. Interestingly, by decoupling the \textit{excess} Balmer absorption from the past averaged star formation rate, the radial gradients in PCA spectral index space show an intriguing feature, starting out positive, flattening and then becoming increasingly negative with time. This should provide an excellent way to measure the age of PSB galaxies, independent of obtaining precise star formation histories, and suggests that the three different gradients observed in MaNGA galaxies in Section \ref{sec:observations} could well be caused by the same physical process caught at different observed times. 

Throughout this paper, we have combined the analysis of two very different spectral index spaces: the equivalent widths of the \ha\ emission line and Balmer absorption lines, and the PCA indices of \citet{wild2007bursty} which focus on the stellar continuum alone. The sensitivity of the \ha\ emission line to residual ongoing star formation, which in our simulations fluctuates wildly dependent on the fluctuating BH feedback strength, provides a natural explanation for the much larger number of PCA-selected PSBs than traditional selection \citep[][]{pawlik2019}. As stated previously, the PCA selection is insensitive to a small amount of residual star formation, which provides an important complementary and more inclusive approach to PSB selection.  However, it is evident from \autoref{fig:observations} that the PCA-selection does not identify all observed PSBs, with 2/3 of the example MaNGA PSBs selected using the traditional method falling outside the PCA PSB region. According to their gradients in PC space, these are likely younger objects, where the Balmer absorption has not yet had time to strengthen and 4000\AA\ break strength to increase sufficiently for PCA selection (i.e. the PCA selection box as used in this paper selects only older PSBs, as noted already previously in \citet{wild2010timing}). The two objects with slightly positive or flat gradients in PC1/2 highlight yet another discrepancy with the simulated mergers: the star formation must have been quenched incredibly rapidly for these to be identified as PSBs through their lack of \ha, and such rapid quenching of \ha\ is never observed in our simulations. This would require an immediate impact of the BH feedback, which as we see from the top panel of \autoref{fig:gal_pro} takes a little while to get going following the coalescence of the BHs. This again suggests an important missing ingredient to the sub-resolution star formation recipes in these simulations. 

Finally, we note that the radial gradients in the spectral indices are caused by a stronger, longer duration starburst occurring in the central regions of the galaxy. The starburst in the outer regions happens co-evally with that in the centre, i.e. not supporting either inside-out nor outside-in growth scenarios. The quenching occurs first in the outer regions, however, supporting outside-in quenching for these extreme galaxies.

\section{Summary}
\label{sec:Summary}

We use Gadget-3 to run a set of binary merger simulations with different black hole feedback models, progenitor galaxies, and orbits. We develop the SEDMorph code to build mock SDSS-like spectra for the simulated galaxy, by combining the star formation history and metallicity of each particle with stellar population synthesis models. We create mock datacubes following the MaNGA observational strategy including a PSF and dithering pattern. The spatial distribution of \ha\ emission, Balmer absorption lines, and stellar continuum shape indices are investigated in the post-merger galaxies, to diagnose recent and ongoing star formation.

A summary of our findings is as follow:
\begin{itemize}
	\item To create mock MaNGA datacubes we found it was necessary to strictly follow the MaNGA observation strategy. Short cuts such as gridding data directly, or convolving with the PSF after data gridding produce significant residuals or larger computational resources compared to a full treatment. 
	\item To completely shut down the star formation in our model PSB galaxies, mechanical AGN feedback is required to expel the gas from the galaxy, while  the more traditional thermal feedback BH model is not sufficiently efficient.
	\item The star formation histories of the merger simulations demonstrate 
	that sharp quenching leading to PSB-like remnants is only achieved in particular circumstances: progenitor galaxies with similar mass, approaching each other in either prograde-prograde or retrograde-prograde orbits. Neither unequal-mass mergers nor retrograde-retrograde mergers lead to quenching that is significant or rapid enough to lead to PSB spectral features.
	\item The traditional PSB selection method, which identifies PSB galaxies via their absence of nebular emission lines as well as strong Balmer absorption lines, is highly sensitive to the complete shut down in star formation, which in turn appears to be sensitive to the accretion rate of the BH shortly before the time of observation. A stellar continuum based method, such as the PCA method used here, is much less sensitive to rapid fluctuations in the SFR of the galaxy and therefore is likely to lead to more complete samples. 
    \item However, two of the three example MaNGA galaxies presented in this paper are not selected by the PCA method as their \ha\ emission has shut off before the Balmer absorption lines are strong enough to be identified cleanly by the PCA. Such a rapid shut off in star formation following the starburst is not found in any of our merger simulations. Combined with the fact that many PSBs are known to have cold gas that is not forming stars efficiently, this points to a missing ingredient in the sub-resolution star formation recipes or the ISM structure employed in these simulations. 
	\item In agreement with previous work, the simulated post-starburst galaxies show a strong radial gradient in the Balmer absorption line strength, with stronger absorption in the inner region, as seen in the majority of local PSB galaxies where the integrated light is dominated by the central PSB region. This appears to be a defining feature of merger-origin PSBs, caused by the inflow of gas to the central regions. 
	\item In PCA space an evolution in the radial gradient becomes apparent, that is masked by traditional methods using nebular emission lines due to their sensitivity to fluctuations in ongoing star formation. This indicates that the range of gradients observed in MaNGA PSB galaxies is simply due to different times of observation rather than different underlying processes. 
	\item Our simulations show that the galaxies undergo a single co-eval burst which was stronger and longer lived in the central regions. This does not support either inside-out nor outside-in growth during the star-formation episode, but rather outside-in quenching. 

\end{itemize}

Clearly much more remains to be understood about the formation of PSB galaxies, with this paper raising questions about the effectiveness of the implementations of BH feedback in the current generation of hydrodynamic simulations. Comparison with cosmological simulations may help us to understand the range of different orbits and gas properties as a function of stellar mass and epoch; however they are still fundamentally limited by resolution as well as the employed sub-resolution star formation and stellar feedback recipes. Observationally, further progress on understanding the causes of both the starburst and final quenching in PSB galaxies may come from combining the analysis of radial gradients in spectral indices with the morphology in deep imaging data, or stellar kinematics from high quality IFU data.

\section*{Acknowledgments}

The authors would like to thank Ena Choi for helpful discussions and guidance on the mechanical BH models, thank Ariel Werle for helping performing PCA with Python,
and Anne-Marie Weijmans for suggestions on Appendix \ref{app:cubecreation}.
YZ acknowledges support of a China Scholarship Council - University of St Andrews Scholarship.  VW and NJ acknowledge support from the European Research Council Starting Grant SEDMorph (P.I. V. Wild). 
NL and PHJ acknowledge support by the European Research Council via ERC Consolidator Grant KETJU (no. 818930)

Funding for the Sloan Digital Sky Survey IV has been provided by the Alfred P. Sloan Foundation, the U.S. Department of Energy Office of Science, and the Participating Institutions. SDSS-IV acknowledges
support and resources from the Center for High-Performance Computing at
the University of Utah. The SDSS web site is www.sdss.org.

SDSS-IV is managed by the Astrophysical Research Consortium for the 
Participating Institutions of the SDSS Collaboration including the 
Brazilian Participation Group, the Carnegie Institution for Science, 
Carnegie Mellon University, the Chilean Participation Group, the French Participation Group, Harvard-Smithsonian Center for Astrophysics, 
Instituto de Astrof\'isica de Canarias, The Johns Hopkins University, Kavli Institute for the Physics and Mathematics of the Universe (IPMU) / 
University of Tokyo, the Korean Participation Group, Lawrence Berkeley National Laboratory, 
Leibniz Institut f\"ur Astrophysik Potsdam (AIP),  
Max-Planck-Institut f\"ur Astronomie (MPIA Heidelberg), 
Max-Planck-Institut f\"ur Astrophysik (MPA Garching), 
Max-Planck-Institut f\"ur Extraterrestrische Physik (MPE), 
National Astronomical Observatories of China, New Mexico State University, 
New York University, University of Notre Dame, 
Observat\'ario Nacional / MCTI, The Ohio State University, 
Pennsylvania State University, Shanghai Astronomical Observatory, 
United Kingdom Participation Group,
Universidad Nacional Aut\'onoma de M\'exico, University of Arizona, 
University of Colorado Boulder, University of Oxford, University of Portsmouth, 
University of Utah, University of Virginia, University of Washington, University of Wisconsin, 
Vanderbilt University, and Yale University.

\section*{Data Availability}
The related simulation data, including snapshots, black hole activity, mock spectra and mock MaNGA datacube are available at: \url{https://doi.org/10.17630/ff244265-6540-494e-af3e-0969fdc5ff24}.
The related MaNGA data are available at the SDSS data base(\url{https://www.sdss.org/dr16/}).
Other data underlying this article are publicly available from the web as listed in the footnote.




\bibliographystyle{mnras}
\bibliography{ref} 



\appendix

\section{Comparing different datacube creation methods}
\label{app:cubecreation}

\begin{figure*}
	\includegraphics[width=1.8\columnwidth]{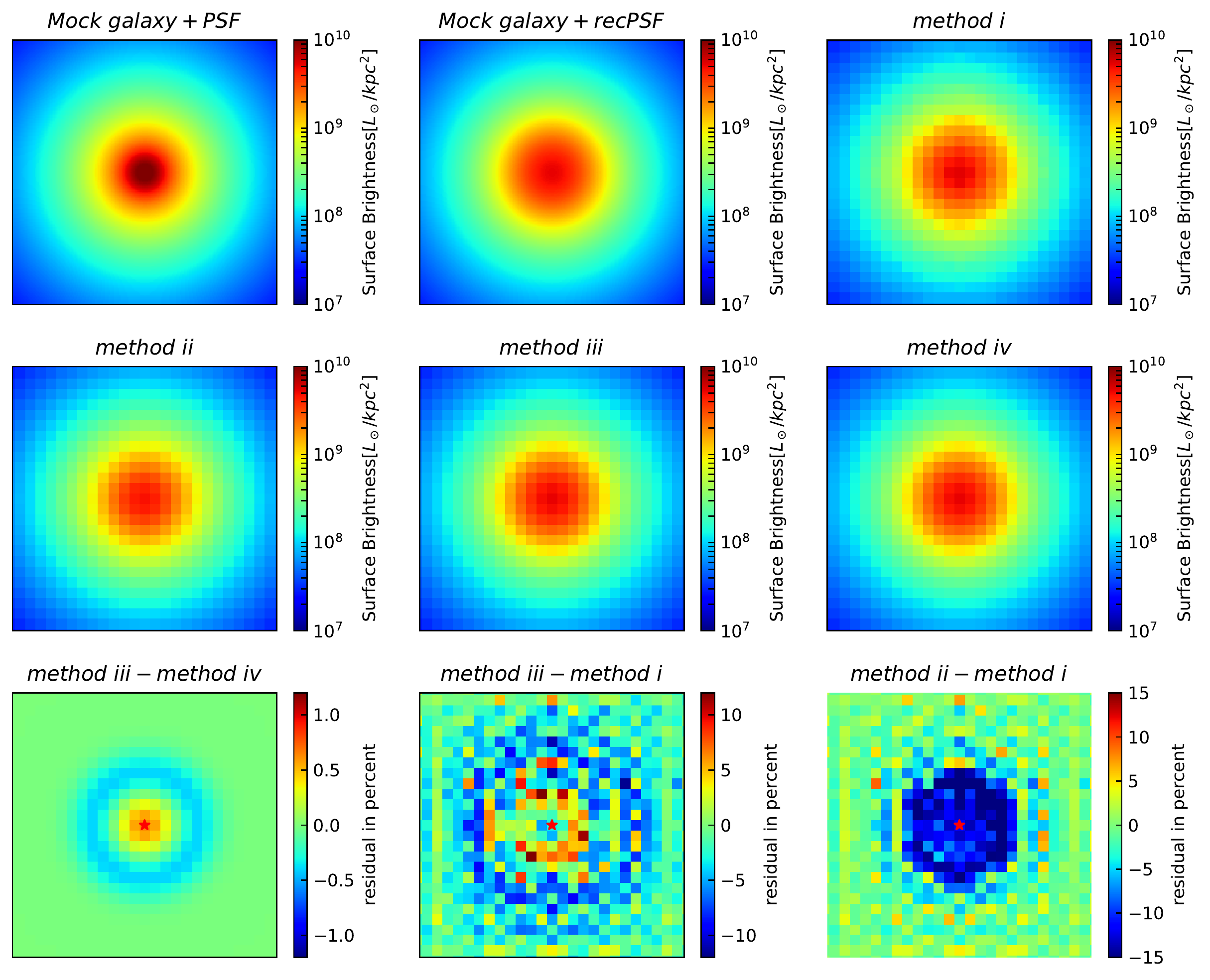}
	\caption{Comparison of different data cube creation methods to turn a mock galaxy into a mock IFU observation. The mock galaxy (top left) has a Sersic light profile with $n=4$ and effective radius of $R_e=1.5\ {\rm kpc}$. All the panels here are $10\times10$\,kpc, displaying the mock galaxy up to $\sim3R_e$. The centre of the galaxies is marked by a red star in the lower panels. The panels are described in detail in the text. The top row shows the mock galaxies with different PSFs as well as the reconstructed image methods that we use in our paper. The middle row shows reconstructed images using the alternative methods described in this appendix. The bottom row shows the comparison between these alternative reconstruction methods and the one we used in our paper.
	}
	\label{fig:compare_cubes}
\end{figure*}

We explored several different algorithms for the creation of a mock datacube. Here we justify why it is necessary to run the full procedure that correctly accounts for the seeing and observed dithering pattern. The methods that we are comparing are:
\begin{enumerate}
    \item use circular fibres, apply the seeing (atmospheric) PSF to the mock galaxy, apply the dither pattern, and subsequently sample the spectra onto a regularly gridded datacube (this is the method described in Section \ref{subsec:datacube});

    \item as method (i), but only convolve the spectra with the seeing after creation of the regularly gridded datacube;

    \item directly sample the galaxy onto a regular square grid, and apply the PSF after creation of the datacube. The PSF for this method consists of the MaNGA reconstructed PSF kernel (\textit{recPSF}), while for the above two methods, the PSF consists of the seeing at the focal plane only.
    \item as method (iii), but  convolve the spectra with the \textit{recPSF} before creation of the regularly gridded datacube. This method is computationally more demanding than method (iii), but also more realistic.
\end{enumerate}
We first create a mock galaxy with a Sersic profile light distribution with $n=4$, effective radius of $R_e=1.5\ {\rm kpc}$, and intensity at the effective radius of  $I_e=10^9\ L_\odot/\rm kpc^2$.
The mock IFUs cover an area of $10\times10\ {\rm kpc^2}$, up to $\sim3R_e$ of the model galaxy. 
The mock galaxy is shown in the top row of \autoref{fig:compare_cubes},
convolved with a typical $g$-band seeing for the SDSS observatory in the top left panel and convolved with the median $g$-band effective reconstructed PSF (\textit{recPSF}) kernel \citep{Yan2016} for the MaNGA survey in the top middle panel.
Note that the \textit{recPSF} is more extended than the seeing PSF because it is the result of the convolution of the Gaussian seeing kernel with the top-hat fibre response function. 
Our full simulated cube is then re-constructed following method (i), which strictly follows the MaNGA observation strategy. 
This image is shown in the top right of \autoref{fig:compare_cubes}. We also reconstruct the image using method (ii) and show the result in the centre-left panel in \autoref{fig:compare_cubes}. Finally, we reconstruct the image using methods  (iii) and (iv), and the resulting images are shown in the centre-middle and centre-right panels, respectively.

In the bottom row of \autoref{fig:compare_cubes} we compare the different methods by showing residuals as indicated. The left panel shows a comparison of method (iii) and (iv), and we conclude that these methods only deviate at the few percent level. The central panel compares method (iii) and (i), which again shows that these methods deliver similar results, with residuals mostly below 5 percent. The benefit of applying the seeing first using the RSS method (method i) is demonstrated in the right panel, with residuals between method (i) and (ii) at the level of 10\%. 
The application of the seeing after the RSS cube construction results in a total flux loss of $\sim$10\%, depending on the exact position of the galaxy centre compared to the IFU centre.

\begin{figure*}

	\includegraphics[width=1.8\columnwidth]{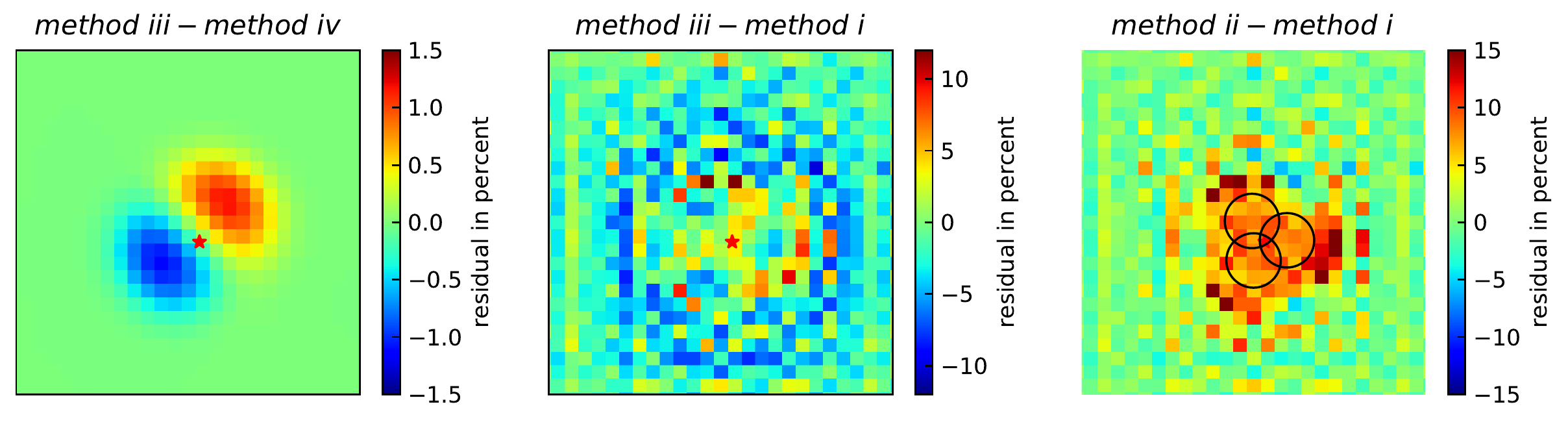}
	\caption{Same as the bottom row of \autoref{fig:compare_cubes} but the centres of the IFUs are slightly offset from the galaxy centre. In this case, the galaxy centre (red star) falls at the position that is always covered by the circular fibres at all three dithered exposures. 
	}
	\label{fig:compare_cubes_tri}
\end{figure*}

In \autoref{fig:compare_cubes_tri} we shift the position of the centres of the IFUs slightly relative to the galaxy centre such that the galaxy centre falls in the region covered by all 3 circular fibres in the dithered MaNGA observations. Method (iii) again produces a similar result compared to the standard RSS method (i). With the RSS method, we find a total flux gain of 5\% when the seeing is applied after (method ii), rather than before the RSS cube creation (method i).

We exclude method (ii) of applying the seeing after the RSS cube as it does not reproduce the MaNGA observations. Using square fibres (methods iii and iv) seems to be acceptable as most pixels have a difference less than 5\%.
However, the square fibre process is more computationally expensive
because it needs to create spectra for every pixel.
For a 127-fibre IFU, the standard RSS method (i) creates only 381 spectra in total (127 fibre $\times$ 3 dithers).The 127-fibre IFU covers a hexagon region with a side width of 12\arcsec. For a pixel size of 0.5\arcsec $\times$ 0.5\arcsec as in MaNGA, this region contains $\sim$1492 pixels.
Therefore, 1492 spectra are required in the square fibre process to reach the same coverage,
which is $\sim$4 times of that in the standard RSS method.
Though the square fibre approaches save the dithering and the data-gridding steps and are faster when creating a single spectra,
it is still 3 times more computationally expensive compared to the RSS approach used in this paper. We therefore settled on method (i) for the datacube reconstruction method used in this paper.

\section{Parameters for the merger simulations}
\label{app:mergertable}
We run 18 equal-mass and 9 unequal-mass merger simulations each with 3 different orbits, giving 27 merger simulations summarised in Table \ref{tab:merger_info}.
\begin{table*}
	
	 \centering
			\caption{Parameters for the merger simulations, where $i$ is the inclination relative to the orbit plane, $\omega$ is the argument of the orbits' pericentre, $r_{\rm sep}$ is the initial separation in kpc and $r_p$ is the pericentre distance in kpc }
			\label{tab:merger_info}
		\begin{tabular}{rcccrrrrrr}
			\multicolumn{10}{c}{\textbf{Equal-mass mergers}}\\
			\hline\hline
			Simulation & First galaxy & Second galaxy & Orbit & 
			$ i_1$ & $i_2$\ & $\omega_1$&$\omega_2$\ &
			$r_{\rm sep}$\ &  $r_p$ \\
			\hline
		2xSa\_00 & Sa & Sa & G00 & $   0^{\circ}$& $  0^{\circ}$ & $   0^{\circ}$ & $   0^{\circ}$ & & \\
		2xSa\_07 & Sa & Sa & G07 & $-109^{\circ}$& $ 71^{\circ}$ & $ -60^{\circ}$ & $ -30^{\circ}$ & & \\
		2xSa\_13 & Sa & Sa & G13 & $-109^{\circ}$& $180^{\circ}$ & $  60^{\circ}$ & $   0^{\circ}$ & & \\
		
		2xSc\_00 & Sc & Sc & G00 & $   0^{\circ}$& $  0^{\circ}$ & $   0^{\circ}$ & $   0^{\circ}$ & & \\
		2xSc\_07 & Sc & Sc & G07 & $-109^{\circ}$& $ 71^{\circ}$ & $ -60^{\circ}$ & $ -30^{\circ}$ & & \\
		2xSc\_13 & Sc & Sc & G13 & $-109^{\circ}$& $180^{\circ}$ & $  60^{\circ}$ & $   0^{\circ}$ & & \\
		
		2xSd\_00 & Sd & Sd & G00 & $   0^{\circ}$& $  0^{\circ}$ & $   0^{\circ}$ & $   0^{\circ}$ & & \\
		2xSd\_07 & Sd & Sd & G07 & $-109^{\circ}$& $ 71^{\circ}$ & $ -60^{\circ}$ & $ -30^{\circ}$ & & \\
		2xSd\_13 & Sd & Sd & G13 & $-109^{\circ}$& $180^{\circ}$ & $  60^{\circ}$ & $   0^{\circ}$ & \multirow{2}{*}{225}& \multirow{2}{*}{7.6}\\

		Sa\_Sc\_00 & Sa & Sc & G00 & $   0^{\circ}$& $  0^{\circ}$ & $   0^{\circ}$ & $   0^{\circ}$  & & \\
		Sa\_Sc\_07 & Sa & Sc & G07 & $-109^{\circ}$& $ 71^{\circ}$ & $ -60^{\circ}$ & $ -30^{\circ}$ & & \\
		Sa\_Sc\_13 & Sa & Sc & G13 & $-109^{\circ}$& $180^{\circ}$ & $  60^{\circ}$ & $   0^{\circ}$ & & \\
		
		Sa\_Sd\_00 & Sa & Sd & G00 & $   0^{\circ}$& $  0^{\circ}$ & $   0^{\circ}$ & $   0^{\circ}$ & & \\
		Sa\_Sd\_07 & Sa & Sd & G07 & $-109^{\circ}$& $ 71^{\circ}$ & $ -60^{\circ}$ & $ -30^{\circ}$ & & \\
		Sa\_Sd\_13 & Sa & Sd & G13 & $-109^{\circ}$& $180^{\circ}$ & $  60^{\circ}$ & $   0^{\circ}$ & & \\
		
		Sc\_Sd\_00 & Sa & Sd & G00 & $   0^{\circ}$& $  0^{\circ}$ & $   0^{\circ}$ & $   0^{\circ}$ & & \\
		Sc\_Sd\_07 & Sa & Sd & G07 & $-109^{\circ}$& $ 71^{\circ}$ & $ -60^{\circ}$ & $ -30^{\circ}$ & & \\
		Sc\_Sd\_13 & Sa & Sd & G13 & $-109^{\circ}$& $180^{\circ}$ & $  60^{\circ}$ & $   0^{\circ}$ & & \\
		\hline
		&&&&&&&&\\
		&&&&&&&&\\
		\multicolumn{10}{c}{\textbf{Unequal-mass  (3:1) mergers}}\\
		\hline\hline
		Simulation & First galaxy & Second galaxy & Orbit& $ i_1$ & $i_2$ & $\omega_1$&$\omega_2$& $r_{\rm sep}$&  $r_p$ \\
		\hline
		Sa\_Scp3\_00 & Sa & Scp3 & G00 & $   0^{\circ}$& $  0^{\circ}$ & $   0^{\circ}$ & $   0^{\circ}$ &  & \\
		Sa\_Scp3\_07 & Sa & Scp3 & G07 & $-109^{\circ}$& $ 71^{\circ}$ & $ -60^{\circ}$ & $ -30^{\circ}$ & & \\
		Sa\_Scp3\_13 & Sa & Scp3 & G13 & $-109^{\circ}$& $180^{\circ}$ & $  60^{\circ}$ & $   0^{\circ}$ & & \\
		
		Sc\_Scp3\_00 & Sc & Scp3 & G00 & $   0^{\circ}$& $  0^{\circ}$ & $   0^{\circ}$ & $   0^{\circ}$ & & \\
		Sc\_Scp3\_07 & Sc & Scp3 & G07 & $-109^{\circ}$& $ 71^{\circ}$ & $ -60^{\circ}$ & $ -30^{\circ}$ & 135.5 & 6.5 \\
		Sc\_Scp3\_13 & Sc & Scp3 & G13 & $-109^{\circ}$& $180^{\circ}$ & $  60^{\circ}$ & $   0^{\circ}$ & & \\
		
		Sd\_Scp3\_00 & Sd & Scp3 & G00 & $   0^{\circ}$& $  0^{\circ}$ & $   0^{\circ}$ & $   0^{\circ}$ & & \\
		Sd\_Scp3\_07 & Sd & Scp3 & G07 & $-109^{\circ}$& $ 71^{\circ}$ & $ -60^{\circ}$ & $ -30^{\circ}$ & & \\
		Sd\_Scp3\_13 & Sd & Scp3 & G13 & $-109^{\circ}$& $180^{\circ}$ & $  60^{\circ}$ & $   0^{\circ}$ & & \\		
		\hline
		\end{tabular}
\end{table*}

\bsp	
\label{lastpage}
\end{document}